\documentclass[twocolumn,showpacs,preprintnumbers]{revtex4}
\usepackage{amssymb}
\usepackage{amsfonts}
\usepackage{amsmath}
\usepackage{graphicx}
\usepackage{dcolumn}
\usepackage{bm}
\usepackage{hyperref}

\input{tcilatex}
\begin{document}

\title{Newtonian gravitational constant measurement. All atomic variables
become extreme when using a source mass consisting of 3 or more parts}
\author{B. Dubetsky}
\affiliation{Independent Researcher, Hallandale Florida 33009, United States}
\email{bdubetsky@gmail.com}
\date{\today }

\begin{abstract}
Atomic interferometry methods used to measure the Newtonian gravitational
constant. To improve the accuracy, one should measure the phase of an atomic
interferometer at extreme values of atomic vertical velocities and
coordinates. Owing to symmetry, the horizontal components of atomic
velocities and coordinates are also extreme. We propose using a source mass
consisting of 3 or more parts, since only in this case one can find such an
arrangement of parts that all atomic variables become extreme. Nonlinear
dependences of the phase on the uncertainties of atomic positions and
velocities near those extreme values required us to modify the expression
for the phase relative standard deviation (RSD). Moreover, taking into
account nonlinear terms in the phase dependence on the atomic coordinates
and velocities leads to a phase shift. In the last experiment to measure the
Newtonian gravitational constant by atomic interferometry, this shift was
not included. We took the shift into account, got a value of $199$ppm for
it, and this leads to a decrease in the value of the Newton constant by $%
0.02\%$. In addition, we showed that at equal sizes of the atomic cloud in
the vertical and horizontal directions, as well as at equal atomic vertical
and transverse temperatures, systematic errors due to the finite size and
temperature of the cloud disappear. The calculation also showed that when
using the 13-ton source mass proposed recently, the measurement accuracy can
reach 17ppm for a source mass consisting of 4 quarters. We assumed that the
source mass consisting of a set of cylinders is used for measurements. We
have obtained a new analytical expression for the gravitational field of a
homogeneous cylinder.
\end{abstract}

\maketitle

\twocolumngrid%

\section{\label{s0}Introduction}

Since its birth about 40 years ago \cite{c1}, the field of atom
interferometry has matured significantly. The current state and prospects in
this area are presented, for example, in the reviews \cite{c1.1} and the
proposals \cite{c1.2,c1.3,c1.4,c1.5,c1.5.1,c1.5.2}.

Among other applications, atom interferometers (AIs) are now used to measure
Newtonian gravity constant $G$ \cite{c12,c2,c3}. Searches for new schemes
and options promise to increase the accuracy of these measurements.
Previously, it was shown \cite{c5} that, in principle, the current
state-of-art in atom interferometry would allow one to measure $G$ with an
accuracy of $200$ppb$.$ In reference \cite{c5} it was assumed that AIs with
the best parameters achieved so far in various experiments \cite{c6,c7,c8,c9}
are used. But even for those parameter values that are currently reached in
the references \cite{c2,c12,c3} one can improve the accuracy of the $G-$%
measurement if one selects the appropriate positions for launching atomic
clouds and the proper parameters of the sources of the gravitational field.
According to \cite{c15}, the main goal here is to reduce the sensitivity of
the AI phase to the initial atomic coordinates. Even more important \cite%
{c5,c10} is the sensitivity to the launching atomic velocities.

\begin{figure}[!t]
\includegraphics[width=8cm]{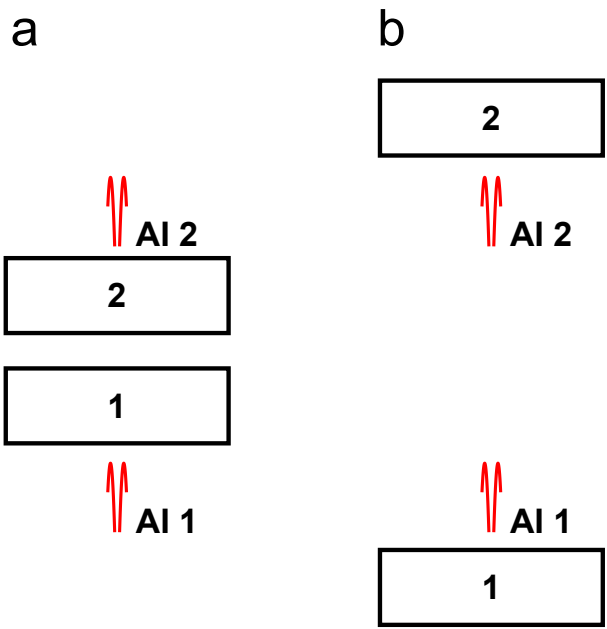}
\caption{Scheme of the $G-$measurement. The atomic gradiometer consists of two AIs.
The phase difference of the AIs is measured in the presence of the
gravitational field of the source mass consisting of two halves 1 and 2.
Measurements are made for two configurations of the source mass `$C$' and `$%
F $'.
(a) In the $C-$configuration, both halves of the source mass are located
between atomic clouds. (b) In the $F-$configuration, on the contrary, the
atomic clouds are located between the halves of the source mass.
Trajectories of atoms are shown in red.}
\label{g0}
\end{figure}

The following procedure was\ used \cite{c12,c3,c5,c10}. The source mass
consists of two halves, which are placed 
in two different configurations C
and F shown in figure \ref{g0}.

We assume the notation `C and F,' which was previously used in article \cite%
{c3}. The atomic gradiometer \cite{c4.1} measures the phase difference of
two atomic interferometers (AIs) $1$ and $2$ 
\begin{equation}
\Delta \phi ^{\left( C,F\right) }=\phi ^{\left( C,F\right) }\left(
z_{1},v_{z_{1}}\right) -\phi ^{\left( C,F\right) }\left(
z_{2},v_{z_{2}}\right) ,  \label{1}
\end{equation}%
where $\phi ^{\left( C,F\right) }\left( z_{j},v_{j}\right) $ is the phase of
AI $j$, in which the atoms are launched vertically from point $\mathbf{x}%
_{j}=\left( 0,0,z_{j}\right) $ at velocity $\mathbf{v}_{j}=\left(
0,0,v_{z_{j}}\right) $. Phase difference (equation (\ref{1})) consists of
two parts, the one that is induced by the gravitational field of the earth
and inertial terms and the other that is associated with the gravitational
field of the source mass. One expects \cite{c2,c12,c3} that the phase double
difference (PDD)%
\begin{equation}
\Delta ^{\left( 2\right) }\phi =\Delta \phi ^{\left( C\right) }-\Delta \phi
^{\left( F\right) }  \label{2}
\end{equation}%
will depend only on the AI phase $\phi _{s}^{\left( C,F\right) }\left(
z_{j},v_{j}\right) $ produced only by the field of source mass, and
therefore can be used to~measure the Newtonian gravitational constant $G$.
Despite the fact that the gravitational field of the earth does not affect
the PDD, the gradient of this field affects \cite{c12,c3} on the accuracy of
the PDD measurement. In the article \cite{c3}, to reduce this influence, the
mutual position of the source mass and atomic clouds are selected so that at
the point of apogee of the atomic trajectories gradients of the earth's
field and the field of the source mass cancel each other. Below in section. %
\ref{s3} we will see that this technique only partially reduces the
influence of the gravitational field of the earth on the accuracy of the $G$
measurement.

A different approach was used in \cite{c12}. In the $C-$configuration, when
all the components of the source mass were located between AIs 1 and 2, for
a given launching velocity 
\begin{equation}
v_{z_{1}}=v_{z_{2}}=v,  \label{n0}
\end{equation}%
varying the position of the atomic cloud 1, one found the point of the local
maximum of the phase $\phi ^{\left( C\right) }\left( z_{1},v\right) $.
Similarly, the cloud of the second interferometer was located at the point
of the local minimum of the phase $\phi ^{\left( C\right) }\left(
z_{2},v\right) $. In the $F-$configuration, one varied the positions of the
source mass halves, $h_{1}$ and $h_{2},$ in order to achieve a situation,
when points $\left( z_{1},v\right) $ and $\left( z_{2},v\right) $ become
respectively the minimum and maximum of the phase $\phi ^{\left( F\right)
}\left( z,v\right) .$ After this procedure, the points $z_{1}$ and $z_{2}$
become extreme in both the C- and $F-$configurations, and therefore they are
extreme for the PDD (equation (\ref{2})). The disadvantage of this approach
is that the atomic velocities $v_{z_{1}}$ and $v_{z_{2}}$ were not varied
and no extreme values were found for them.

To overcome this difficulty, we propose to divide the source mass into a
larger number of parts, the position of each of which can vary
independently. In the $C-$configuration, when all the parts are put
together, under a sufficiently strong gravitational field, one can still
find the points of local maximum and minimum $\left( z_{1},v_{z_{1}}\right) $
and $\left( z_{2},v_{z_{2}}\right) $ and place atomic clouds in those
points. Our goal is that in the $F-$configuration the same points still
remain extreme, i.e. they satisfy a system of 4 equations%
\begin{equation}
\partial _{z_{j}}\phi ^{\left( F\right) }\left( z_{j},v_{z_{j}}\right)
=\partial _{v_{z_{j}}}\phi ^{\left( F\right) }\left( z_{j},v_{z_{j}}\right)
=0,  \label{n1}
\end{equation}%
where $j=1$ or $2.$ Although the points are given, the phase $\phi ^{\left(
F\right) }\left( z,v\right) $ is a function of the coordinates of the source
mass parts, such as their location along the vertical axis, $h_{1},\ldots
h_{n}$, where $n$ is the number of parts. Then equation (\ref{n1}) should be
considered as a system of equations for $\left( h_{1},\ldots h_{n}\right) $.
Since the number of equations and the number of variables must coincide, we
conclude that the source mass must consist of four parts. However,
calculations have shown that the extreme values of the velocities of both
atomic clouds in the $C-$configuration coincide. Therefore, it is sufficient
to divide the source mass into 3 parts to make extreme all atomic variables
in the both configuration.

To test the feasibility of our proposals, we compared the error budgets in
our case and in article \cite{c3}. In precision gravity experiments, one
calculates or measures the standard deviation (SD) $\sigma $ of the response 
$f$ (such as the AI phase or phase difference) using the expression

\begin{equation}
\sigma \left( f\right) =\left( \sum_{m=1}^{n}\sigma _{m}^{2}\right) ^{1/2},
\label{3}
\end{equation}%
where $n$ is the number of variables $\left\{ q_{1},\ldots q_{n}\right\} $,
included in the error budget, $\sigma _{m}=\left\vert \partial f/\partial
q_{m}\right\vert \sigma \left( q_{m}\right) ,$ and $\sigma \left(
q_{m}\right) $ is a SD of small uncertainty in the variable $q_{m}.$ We
assume that variables $\left\{ q_{1},\ldots q_{n}\right\} $ are
statistically independent. See examples of such budgets in \cite%
{c2,c12,c3,c16,c15}. The situation changes when one considers uncertainties
near the extreme points $\left\{ \mathbf{x}_{m},\mathbf{v}_{m}\right\} $ and
the signal's uncertainty becomes a quadratic function of the uncertainties
of the atomic position and velocity $\left\{ \delta \mathbf{x}_{m},\delta 
\mathbf{v}_{m}\right\} .$ There are several examples in which measurements
were carried out (or proposed to be carried out) near extreme points.
Extreme atomic coordinates were selected in the experiments \cite{c12}.
Extreme atomic coordinates and velocities were found in the articles \cite%
{c5,c10}. The difficulties of using extreme points are noted in the article 
\cite{c15}, where an alternative approach was proposed, based on the
elimination of the dependence of the AI phase on the atomic position and
velocity proposed in \cite{c17}. However, even in this case, one eliminates
only the dependence on the vertical coordinates and velocities, while the
transverse coordinates $\left\{ x_{m},y_{m}\right\} =\left\{ 0,0\right\} $
and velocities $\left\{ v_{x_{m}},v_{y_{m}}\right\} =\left\{ 0,0\right\} $
remain extreme. This is because the vertical component of the gravitational
field of the hollow cylinder $\delta g_{3}\left( \mathbf{x}\right) $ is
axially symmetric, and the expansion of both the field and the field
gradient in transverse coordinates begins with quadratic terms. Transverse
velocities and coordinates were also extreme in experiment \cite{c3}. Since
for extreme variables $\partial f/\partial q_{m}=0,$ one sees that in all
the cases listed above \cite{c3,c5,c10,c12,c15}, the use of the expression
(equation (\ref{3})) is unjustified. Revision of this expression is
required. Moreover, the quadratic dependence on the uncertainties $\left\{
\delta \mathbf{x}_{m},\delta \mathbf{v}_{m}\right\} $ leads to a shift in
the signal \cite{c18}. Here, we carried out this revision and expressed both
the SD and the shift of the PDD (equation (\ref{2})) in terms of the first
and second derivatives of the phases $\phi ^{\left( C,F\right) }$ to find
contributions to an error budget from both extreme and non-extreme variables.

Recently, we performed \cite{c5,c10} calculations, determined the optimal
geometry of the gravitational field, positions and velocities of atomic
clouds for the source mass of a cuboid shape. The choice of this shape is
convenient for calculations since one has an analytical expression for the
potential of the cuboid \cite{c13}. Despite this, it is preferable to use
the source mass in a cylindrical shape to perform high-precision
measurements of $G$ \cite{c14}. Cylindrical source masses were used to
measure $G$ in \cite{c12,c3}. The hollow cylinder source mass has been
proposed to achieve an accuracy of 10ppm \cite{c15}. The analytical
expression for the gravitational field along the $z-$axis of the hollow
cylinder was explored \cite{c15}, but outside this axis, the potential
expansion into spherical harmonics was used \cite{c12,c3}. Fast converging
power series for the potential and axial component of the cylinder's
gravitational field were obtained in reference \cite{c18.1}. Analytical
expressions for the field of the cylinders have been derived in the articles 
\cite{c15.1,c15.2}. Alternatively, the technique for calculating the
gravitational field without calculating the gravitational potential was
proposed in the book \cite{c14}, but the final expression for the cylinder
field is given in \cite{c14} without derivation. Following technique \cite%
{c14}, we calculated the field and arrived at expressions (equations (\ref%
{a16}) and (\ref{a20})). Our expressions do not coincide with those given in 
\cite{c14,c15.1,c15.2}. Both the derivations and final results are presented
in this article. Following the derivations in the articles \cite{c15.1,c15.2}%
, we are going to find out analytically the reason of the discrepancies
between different expressions and publish it elsewhere.

The article is arranged as follows.\ SD and shift are obtained in the next
section, where terms nonlinear in the atomic variables variations have been
included in section \ref{s2.1}. Section \ref{s2.2} is devoted to the AI
phase and phase derivatives calculations. PDD and error budget for the
scheme chosen in the article \cite{c3} are considered in the section. \ref%
{s3}. The 3-part source mass is considered in the section \ref{s4}. In the
section \ref{s4.1}, it is shown that for the same total weight of the source
mass, dividing it into 3 equal parts allows one to find a scheme in which
all atomic variables become extreme, and the calculation of the PDD and
error budget for this scheme are carried out. In the Sec. \ref{s4.2}, a
calculation was made for parameters suggested by G. Rossi \cite{c15}. The
conclusions are given in Sec. \ref{s5}. Details of the numerical
calculations and a derivation of the formula for the gravitational field of
the cylinder are presented in the Appendixes.

\section{Error budget near extreme atomic variables}

\subsection{\label{s2.1}SD and shift.}

Let us consider the variation of the double difference (\ref{2})%
\begin{gather}
\delta \Delta ^{\left( 2\right) }\phi \left[ \delta \mathbf{x}_{1C},\delta 
\mathbf{v}_{1C},\delta \mathbf{x}_{2C},\delta \mathbf{v}_{2C};\delta \mathbf{%
x}_{1F},\delta \mathbf{v}_{1F};\delta \mathbf{x}_{2F},\delta \mathbf{v}_{2F}%
\right]  \notag \\
=\delta \phi ^{\left( C\right) }\left[ \delta \mathbf{x}_{1C},\delta \mathbf{%
v}_{1C}\right] -\delta \phi ^{\left( C\right) }\left[ \delta \mathbf{x}%
_{2C},\delta \mathbf{v}_{2C}\right]  \notag \\
-\left[ \delta \phi ^{\left( F\right) }\left( \delta \mathbf{x}_{1F},\delta 
\mathbf{v}_{1F}\right) -\delta \phi ^{\left( F\right) }\left( \delta \mathbf{%
x}_{2F},\delta \mathbf{v}_{2F}\right) \right] ,  \label{4}
\end{gather}%
where $\left\{ \delta \mathbf{x}_{jI},\delta \mathbf{v}_{jI}\right\} $ is
the uncertainty of the launching position and velocity of the cloud $j$ $%
\left( j=1\text{ or }2\right) $ for the source mass configuration $I$ $%
\left( I=C\text{ or }F\right) ,~\delta \phi ^{\left( I\right) }\left( \delta 
\mathbf{x}_{jI},\delta \mathbf{v}_{jI}\right) $ is the variation of the AI $%
j $ phase, produced when the source mass gravity field is in the $I-$%
configuration. For the shift $s$ and standard deviation $\sigma $ defined as 
\begin{subequations}
\label{5}
\begin{gather}
s\left( \Delta ^{\left( 2\right) }\phi \right) =  \notag \\
\left\langle \delta \Delta ^{\left( 2\right) }\phi \left[ \delta \mathbf{x}%
_{1C},\delta \mathbf{v}_{1C},\delta \mathbf{x}_{2C},\delta \mathbf{v}%
_{2C};\delta \mathbf{x}_{1F},\delta \mathbf{v}_{1F};\delta \mathbf{x}%
_{2F},\delta \mathbf{v}_{2F}\right] \right\rangle ,  \label{5a} \\
\sigma \left( \Delta _{s}^{\left( 2\right) }\phi \right) =\left\{
\left\langle \left[ \delta \Delta ^{\left( 2\right) }\phi \left( \delta 
\mathbf{x}_{1C},\delta \mathbf{v}_{1C},\delta \mathbf{x}_{2C},\delta \mathbf{%
v}_{2C};\right. \right. \right. \right.  \notag \\
\left. \left. \left. \left. \delta \mathbf{x}_{1F},\delta \mathbf{v}%
_{1F};\delta \mathbf{x}_{2F},\delta \mathbf{v}_{2F}\right) \right]
^{2}\right\rangle -s^{2}\left( \Delta ^{\left( 2\right) }\phi \right)
\right\} ^{1/2}  \label{5bb}
\end{gather}%
one finds 
\end{subequations}
\begin{subequations}
\label{6}
\begin{gather}
s\left( \Delta ^{\left( 2\right) }\phi \right) =s\left[ \phi ^{\left(
C\right) }\left( \delta \mathbf{x}_{1C},\delta \mathbf{v}_{1C}\right) \right]
-s\left[ \phi ^{\left( C\right) }\left( \delta \mathbf{x}_{2C},\delta 
\mathbf{v}_{2C}\right) \right]  \notag \\
-s\left[ \phi ^{\left( F\right) }\left( \delta \mathbf{x}_{1F},\delta 
\mathbf{v}_{1F}\right) \right] +s\left[ \phi ^{\left( F\right) }\left(
\delta \mathbf{x}_{2F},\delta \mathbf{v}_{2F}\right) \right] ,  \label{6a} \\
\sigma \left( \Delta ^{\left( 2\right) }\phi \right) =\left\{
\dsum_{I=C,F}\dsum_{j=1,2}\sigma ^{2}\left[ \phi ^{\left( I\right) }\left(
\delta \mathbf{x}_{jI},\delta \mathbf{v}_{jI}\right) \right] \right\} ^{1/2},
\label{6b} \\
s\left[ \phi ^{\left( I\right) }\left( \delta \mathbf{x}_{jI},\delta \mathbf{%
v}_{jI}\right) \right] =\left\langle \delta \phi ^{\left( I\right) }\left(
\delta \mathbf{x}_{jI},\delta \mathbf{v}_{jI}\right) \right\rangle ,
\label{6c} \\
\sigma \left[ \phi ^{\left( I\right) }\left( \delta \mathbf{x}_{jI},\delta 
\mathbf{v}_{jI}\right) \right] =\left\{ \left\langle \left[ \delta \phi
^{\left( I\right) }\left( \delta \mathbf{x}_{jI},\delta \mathbf{v}%
_{jI}\right) \right] ^{2}\right\rangle \right.  \notag \\
\left. -\left\langle \delta \phi ^{\left( I\right) }\left( \delta \mathbf{x}%
_{jI},\delta \mathbf{v}_{jI}\right) \right\rangle ^{2}\right\} ^{1/2}
\label{6d}
\end{gather}%
One sees that the problem is reduced to the calculation of the shift $s$ and
SD $\sigma $ of a variation $\delta \phi \left[ \delta \mathbf{x},\delta 
\mathbf{v}\right] .$ The phase of the given AI at the given configuration of
the source mass comprises two parts 
\end{subequations}
\begin{equation}
\phi \left( \mathbf{x},\mathbf{v}\right) =\phi _{E}\left( \mathbf{x},\mathbf{%
v}\right) +\phi _{s}\left( \mathbf{x},\mathbf{v}\right) ,  \label{2.3}
\end{equation}%
where for the phase induced by the earth's field, under some simplifying
assumptions (see, for example, \cite{c19}), one gets 
\begin{eqnarray}
\phi _{E}^{\left( I\right) }\left( \mathbf{x},\mathbf{v}\right) &=&\mathbf{%
k\cdot g}T^{2}+\mathbf{k\cdot }\Gamma _{E}T^{2}\left[ \mathbf{x}+\mathbf{v}%
\left( T+T_{1}\right) \right]  \notag \\
&&+\mathbf{k\cdot }\Gamma _{E}\mathbf{g}T^{2}\left( \dfrac{7}{12}%
T^{2}+TT_{1}+\dfrac{1}{2}T_{1}^{2}\right) ,  \label{2.4}
\end{eqnarray}%
where $T_{1}$ is the time delay between the moment the atoms are launched
and the 1st Raman pulse. For the vertical wave vector $\mathbf{k=}\left(
0,0,k\right) ,$ expanding equation (\ref{2.3}).to the second order terms one
gets 
\begin{gather}
\delta \phi \left( \delta \mathbf{x},\delta \mathbf{v}\right) =\left( \tilde{%
\gamma}_{xm}+\dfrac{\partial \phi _{s}}{\partial x_{m}}\right) \delta
x_{m}+\left( \tilde{\gamma}_{vm}+\dfrac{\partial \phi _{s}}{\partial v_{m}}%
\right) \delta v_{m}  \notag \\
+\dfrac{1}{2}\dfrac{\partial ^{2}\phi _{s}}{\partial x_{m}\partial x_{n}}%
\delta x_{m}\delta x_{n}+\dfrac{1}{2}\dfrac{\partial ^{2}\phi _{s}}{\partial
v_{m}\partial v_{n}}\delta v_{m}\delta v_{n}  \notag \\
+\dfrac{\partial ^{2}\phi _{s}}{\partial x_{m}\partial v_{n}}\delta
x_{m}\delta v_{n},  \label{7}
\end{gather}%
where 
\begin{subequations}
\label{7.1}
\begin{eqnarray}
\tilde{\gamma}_{xm} &=&k\Gamma _{E3m}T^{2};  \label{7.1a} \\
\tilde{\gamma}_{vm} &=&\left( T+T_{1}\right) \tilde{\gamma}_{xm},
\label{7.1b}
\end{eqnarray}%
where $\Gamma _{E3m}$ is the $3m-$component of the earth's field
gravity-gradient tensor. A summation convention implicit in equation (\ref{7}%
) will be used in all subsequent equations. Repeated indices and symbols
appearing on the right-hand-side of an equation are to be summed over,
unless they also appear on the left-hand-side of that equation. Let assume
that the distribution functions of the uncertainties are sufficiently
symmetric, and all odd moments are equal $0.$ The moments of the second and
fourth orders are given by 
\end{subequations}
\begin{subequations}
\label{8}
\begin{gather}
\left\langle \delta q_{m}\delta q_{n}\right\rangle =\delta _{mn}\sigma
^{2}\left( q_{m}\right) ,  \label{8a} \\
\left\langle \delta q_{m}\delta q_{n}\delta q_{m^{\prime }}\delta
q_{n^{\prime }}\right\rangle =\delta _{mn}\delta _{m^{\prime }n^{\prime
}}\sigma ^{2}\left( q_{m}\right) \sigma ^{2}\left( q_{m^{\prime }}\right) 
\notag \\
+\left( \delta _{mm^{\prime }}\delta _{nn^{\prime }}+\delta _{mn^{\prime
}}\delta _{nm^{\prime }}\right) \sigma ^{2}\left( q_{m}\right) \sigma
^{2}\left( q_{n}\right)  \notag \\
+\delta _{mn}\delta _{mm^{\prime }}\delta _{mn^{\prime }}\kappa \left(
q_{m}\right) \sigma ^{4}\left( q_{m}\right) ,  \label{8b}
\end{gather}%
where $q_{i}$ is either a position $x_{i}$ or a velocity $v_{i},$ $\sigma
\left( q_{i}\right) $ is SD of the uncertainty $\delta q_{i},$ $\delta _{mn}$
is Kronecker symbol, $\kappa \left( q_{m}\right) $ is a cumulant of the
given uncertainty $\delta q_{m}$, defined as 
\end{subequations}
\begin{equation}
\kappa \left( q_{m}\right) =\dfrac{\left\langle \delta
q_{m}^{4}\right\rangle }{\sigma ^{4}\left( q_{m}\right) }-3.  \label{9}
\end{equation}%
Let assume also that uncertainties of the launching velocities and positions
are statistically independent,%
\begin{equation}
\left\langle \delta x_{m}\delta v_{n}\right\rangle =0.  \label{9.1}
\end{equation}%
Using the moments (equations (\ref{8}) and (\ref{9.1})) one arrives at the
following expressions for the SD and shift 
\begin{subequations}
\label{10}
\begin{align}
\sigma \left[ \phi \left( \delta \mathbf{x},\delta \mathbf{v}\right) \right]
& =\left\{ \left( \tilde{\gamma}_{xm}+\dfrac{\partial \phi _{s}}{\partial
x_{m}}\right) ^{2}\sigma ^{2}\left( x_{m}\right) \right.  \notag \\
& +\left( \tilde{\gamma}_{vm}+\dfrac{\partial \phi _{s}}{\partial v_{m}}%
\right) ^{2}\sigma ^{2}\left( v_{m}\right)  \notag \\
& +\dfrac{1}{2}\left[ \left( \dfrac{\partial ^{2}\phi }{\partial
x_{m}\partial x_{n}}\right) ^{2}\sigma ^{2}\left( x_{m}\right) \sigma
^{2}\left( x_{n}\right) \right.  \notag \\
& \left. +\left( \dfrac{\partial ^{2}\phi }{\partial v_{m}\partial v_{n}}%
\right) ^{2}\sigma ^{2}\left( v_{m}\right) \sigma ^{2}\left( v_{n}\right) %
\right]  \notag \\
& +\left( \dfrac{\partial ^{2}\phi }{\partial x_{m}\partial v_{n}}\right)
^{2}\sigma ^{2}\left( x_{m}\right) \sigma ^{2}\left( v_{n}\right)  \notag \\
& +\dfrac{1}{4}\left[ \left( \dfrac{\partial ^{2}\phi }{\partial x_{m}^{2}}%
\right) ^{2}\kappa \left( x_{m}\right) \sigma ^{4}\left( x_{m}\right) \right.
\notag \\
& \left. \left. +\left( \dfrac{\partial ^{2}\phi }{\partial v_{m}^{2}}%
\right) ^{2}\kappa \left( v_{m}\right) \sigma ^{4}\left( v_{m}\right) \right]
\right\} ^{1/2},  \label{10a} \\
s\left[ \phi \left( \delta \mathbf{x},\delta \mathbf{v}\right) \right] & =%
\dfrac{1}{2}\left( \dfrac{\partial ^{2}\phi }{\partial x_{m}^{2}}\sigma
^{2}\left( x_{m}\right) +\dfrac{\partial ^{2}\phi }{\partial v_{m}^{2}}%
\sigma ^{2}\left( v_{m}\right) \right) ,  \label{10b}
\end{align}%
One sees that, even for the symmetric uncertainties distribution, the
knowledge of the uncertainties' SDs is not sufficient. One has to know also
uncertainties' cumulants (equation (\ref{9})). The exclusion here is
Gaussian distributions, for which the cumulants 
\end{subequations}
\begin{equation}
\kappa \left( x_{m}\right) =\kappa \left( v_{m}\right) =0.  \label{10.1}
\end{equation}%
Further calculations will be performed only for these distributions.

For the each case considered below we are going to calculate the double
difference (equation (\ref{2}))$\ $and relative contributions to the SD
(equation (\ref{10a})) and shift (equation (\ref{10b})) from the each of two
atom clouds at the each of two source mass configurations.

\subsection{\label{s2.2}The phase and phase derivatives of the atom
interferometer}

To calculate the phase $\phi _{s}^{\left( I\right) }$ produced by the
gravitational field of the source mass, we use the results obtained in the
article \cite{c20}.\ It is necessary to distinguish three contributions to
the phase, classical, quantum, and Q-term [see equations ((62c), (64),
(60c)), ((62d), (71), (60c)) and (89) in \cite{c20} for these three terms].
The quantum term arises from the quantum kicks $\pm \hbar \mathbf{k}$ of the
atomic momentum when interacting with the Raman pulse, while the Q-term is
due to quantum corrections to the free evolution of the density matrix in
the Wigner representation in the space between Raman pulses.

For Q-term an estimate was obtained 
\begin{equation}
\dfrac{\phi _{Q}}{\phi _{s}^{\left( I\right) }}\thicksim \dfrac{1}{24}\left( 
\dfrac{\hbar kT}{LM_{a}}\right) ^{2},  \label{11}
\end{equation}%
where $M_{a}$ is the atom mass, $L$ is the characteristic distance over
which the gravitational potential of the test mass changes. For $^{\text{87}%
} $Rb, at $L>0.3$m, the relative weight of the Q-term does not exceed 2ppb,
and we neglect it. For the remaining terms and the vertical effective wave
vector, $\mathbf{k}=\left\{ 0,0,k\right\} ,$ one gets%
\begin{equation}
\phi _{s}^{\left( I\right) }\left( \mathbf{x},\mathbf{v}\right)
=k\int_{0}^{T}dt\left[ \left( T-t\right) \delta g_{3}\left( \mathbf{a}\left(
T+t\right) \right) +t\delta g_{3}\left( \mathbf{a}\left( t\right) \right) %
\right] ,  \label{12}
\end{equation}%
where%
\begin{equation}
\mathbf{a}\left( t\right) =\mathbf{x}+\mathbf{v}\left( T_{1}+t\right) +%
\dfrac{1}{2}\mathbf{g}\left( T_{1}+t\right) ^{2}+\mathbf{v}_{r}t,
\label{12.1}
\end{equation}%
the recoil velocity is given by%
\begin{equation}
\mathbf{v}_{r}=\hbar \mathbf{k/}2M_{a},  \label{12.2}
\end{equation}%
$\delta g_{3}\left( \mathbf{x}\right) $ is the vertical component of the
gravitational field of the source mass. The derivatives of this phase of the
first and second order are given by 
\begin{subequations}
\label{13}
\begin{gather}
\dfrac{\partial \phi _{s}^{\left( I\right) }\left( \mathbf{x},\mathbf{v}%
\right) }{\partial x_{m}}=k\int_{0}^{T}dt\left[ \left( T-t\right) \Gamma
_{s3m}\left( \mathbf{a}\left( T+t\right) \right) \right.  \notag \\
\left. +t\Gamma _{s3m}\left( \mathbf{a}\left( t\right) \right) \right] ,
\label{13a} \\
\dfrac{\partial \phi _{s}^{\left( I\right) }\left( \mathbf{x},\mathbf{v}%
\right) }{\partial v_{m}}=k\int_{0}^{T}dt\left[ \left( T-t\right) \left(
T_{1}+T+t\right) \right.  \notag \\
\times \left. \Gamma _{s3m}\left( \mathbf{a}\left( T+t\right) \right)
+t\left( T_{1}+t\right) \Gamma _{s3m}\left( \mathbf{a}\left( t\right)
\right) \right] ,  \label{13b} \\
\dfrac{\partial ^{2}\phi _{s}^{\left( I\right) }\left( \mathbf{x},\mathbf{v}%
\right) }{\partial x_{m}\partial x_{n}}=k\int_{0}^{T}dt\left[ \left(
T-t\right) \chi _{s3mn}\left( \mathbf{a}\left( T+t\right) \right) \right. 
\notag \\
\left. +t\chi _{s3mn}\left( \mathbf{a}\left( t\right) \right) \right] ,
\label{13c} \\
\dfrac{\partial ^{2}\phi _{s}^{\left( I\right) }\left( \mathbf{x},\mathbf{v}%
\right) }{\partial x_{m}\partial v_{n}}=k\int_{0}^{T}dt\left[ \left(
T-t\right) \left( T_{1}+T+t\right) \right.  \notag \\
\times \left. \chi _{s3mn}\left( \mathbf{a}\left( T+t\right) \right)
+t\left( T_{1}+t\right) \chi _{s3mn}\left( \mathbf{a}\left( t\right) \right) %
\right] ,  \label{13d} \\
\dfrac{\partial ^{2}\phi _{s}^{\left( I\right) }\left( \mathbf{x},\mathbf{v}%
\right) }{\partial v_{m}\partial v_{n}}=k\int_{0}^{T}dt\left[ \left(
T-t\right) \left( T_{1}+T+t\right) ^{2}\right.  \notag \\
\times \left. \chi _{s3mn}\left( \mathbf{a}\left( T+t\right) \right)
+t\left( T_{1}+t\right) ^{2}\chi _{s3mn}\left( \mathbf{a}\left( t\right)
\right) \right] ,  \label{13e}
\end{gather}%
where $\Gamma _{s3m}\left( \mathbf{x}\right) =\dfrac{\partial \delta
g_{3}\left( \mathbf{x}\right) }{\partial x_{m}}$ is the $3m-$component of
the gravity-gradient tensor of the source mass field, and 
\end{subequations}
\begin{equation}
\chi _{s3mn}\left( \mathbf{x}\right) =\dfrac{\partial ^{2}\delta g_{3}\left( 
\mathbf{x}\right) }{\partial x_{m}\partial x_{n}}  \label{13.1}
\end{equation}%
is the $3mn-$component of the curvature tensor of this field.

In these expressions, any points on the atomic trajectory can be chosen as
atomic variables, i.e. the equations (\ref{2.4}), (\ref{12}), and (\ref{13})
are invariant under replacement 
\begin{subequations}
\label{13.2}
\begin{eqnarray}
\left\{ \vec{x},\vec{v}\right\} &\rightarrow &\left\{ \vec{x}^{\prime },\vec{%
v}^{\prime }\right\} =\left\{ \vec{x}+\vec{v}T^{\prime }+\dfrac{1}{2}\vec{g}%
T^{\prime 2},\vec{v}+\vec{g}T^{\prime }\right\} ,  \label{13.2a} \\
T_{1} &\rightarrow &T_{1}-T^{\prime }.  \label{13.2b}
\end{eqnarray}%
This freedom of choice allowed in reference \cite{c3} the apogee of the
atomic trajectory in the lower interferometer to be used as a reference
point. However, the situation changes when one builds an error budget. If
the atomic coordinates and velocities are statistically independent at the
launch point, then this independence is violated at all other points, since
at them $\delta \vec{x}^{\prime }=\delta \vec{x}+T^{\prime }\delta \vec{v}$,
and, consequently, instead of equation (\ref{9.1}) one gets 
\end{subequations}
\begin{equation}
\left\langle \delta x_{m}^{\prime }\delta v_{n}^{\prime }\right\rangle
=\delta _{mn}\sigma ^{2}\left( v_{m}\right) T^{\prime }\not=0.  \label{13.3}
\end{equation}%
Since statistical independence is necessary for calculating the error budget
using both the generally accepted equations (\ref{3}) and. (\ref{10a}), the
launch point should be preferred for calculations. We did not find
information about the delay time $T_{1}$ between the moment of atom launch
and the first Raman pulse. An example of calculating the AI phase at $%
T_{1}\not=0$ can be found in the article \cite{c10}. In this article, all
calculations are made under the assumption that%
\begin{equation}
T_{1}\ll T.  \label{13.4}
\end{equation}

\section{\label{s3}Corrections to the former error budget.}

\begin{figure}[!t]
\includegraphics[width=8cm]{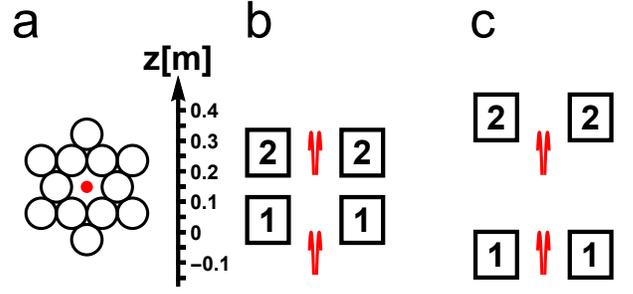}
\caption{The mutual positioning of the
source mass halves 1 and 2, and atomic clouds. Top view (a), cross-sections $%
x=0$ for $C-$configuration (b) and $F-$configuration (c).}
\label{g1}
\end{figure}

We applied the formula for the cylinder field (equation (\ref{a16})) to
calculate the phases produced by different sets of cylinders. In this
section, we consider the field geometry chosen in the article \cite{c3}, see
figure \ref{g1}.

Two halves of the source mass, each including 12 tungsten alloy cylinders,
move in a vertical direction from $C-$configuration to $F-$configuration, in
each of which one measures the phase difference of the first order equation (%
\ref{1}), and then PDD equation (\ref{2}). The following system parameters
are important for calculation: cylinder density $\rho =18263$kg/m%
${{}^3}$%
, cylinder radius and height $R=0.04995$m and $h=0.15011$m, Newtonian
gravitational constant $G=6.67430\cdot 10^{-11}$kg$^{\text{-1}}$m$^{\text{3}%
} $s$^{\text{-2}}$ \cite{c21}, the earth's gravitational field $g=9.80492$m s%
$^{-\text{2}}$ \cite{c22}, the delay between impulses $T=160$ms, the
effective wave vector $k=1.61058\cdot 10^{7}$m$^{\text{-1}}$, the mass of
the $^{\text{87}}$Rb $M_{a}=86.9092$a.u. \cite{c23}, atomic velocity at the
moment of the first impulse action $v=1.62762$m s$^{-1}$ \cite{c4}. With
respect to the apogee of the atomic trajectory in the lower interferometer,
the $z-$coordinates of the centers of the halves of the source mass are
equal to $0.04$m and $0.261$m in the $C-$configuration and $-0.074$m and $%
0.377$m in the $F-$configuration, $z-$coordinate of the atomic trajectory
apogee in the upper interferometer is equal to $0.328$m (see figures \ref{g1}
(b) and (c)). Using equation (\ref{12}) we got for PDD 
\begin{equation}
\Delta ^{\left( 2\right) }\phi =0.530552\text{rad,}  \label{14}
\end{equation}%
which is less than the value obtained in the article \cite{c3}, by 3.2\%.
The difference seems to be related to the fact that in our calculations, the
contributions from platforms and other sources of gravity were not taken
into account. Details of the calculations of the error budget one can find
in Appendix \ref{m1}. For SDs achieved in \cite{c3} 
\begin{subequations}
\label{17}
\begin{eqnarray}
\sigma \left( x_{jI}\right) &=&\sigma \left( y_{jI}\right) =10^{-3}\text{m},
\label{17a} \\
\sigma \left( z_{jI}\right) &=&10^{-4}\text{m},  \label{17b} \\
\sigma \left( v_{xjI}\right) &=&\sigma \left( v_{yjI}\right) =6\cdot 10^{-3}%
\text{m s}^{-1},  \label{17c} \\
\sigma \left( v_{zjI}\right) &=&3\cdot 10^{-3}\text{m s}^{-1}\text{,}
\label{17d}
\end{eqnarray}%
using equation (\ref{S.1}) one arrives to the RSD and the shift 
\end{subequations}
\begin{subequations}
\label{18}
\begin{gather}
\sigma \left( \delta \Delta _{s}^{\left( 2\right) }\phi \right) =275\text{ppm%
}\left[ 1+6.14\cdot 10^{13}\left( \Gamma _{E31}^{2}+\Gamma _{E32}^{2}\right) %
\right] ^{1/2}\text{,}  \label{18a} \\
s\left( \delta \Delta _{s}^{\left( 2\right) }\phi \right) =199\text{ppm.}
\label{18b}
\end{gather}%
The non-diagonal matrix elements of the gradient tensor of the earth's field
consist of three contributions arising from the fact that the Geoid is not
spherical, from the rotation of the earth, and from the anomalous part of
the field. The first two contributions were taken into account exactly \cite%
{c1.7}, and they are $3$ orders of magnitude smaller than the diagonal
element $\Gamma _{E33}$. We failed to find any information about the
anomalous part of the earth's gravitational field. However, it is seen that
the non-diagonal elements of the tensor can be neglected with an accuracy of
not more than 10\% if 
\end{subequations}
\begin{equation}
\sqrt{\Gamma _{E31}^{2}+\Gamma _{E32}^{2}}<58.5\text{E}.  \label{18.1}
\end{equation}

\section{\label{s4}Source mass consisting of 3 parts.}

\subsection{\label{s4.1}Using current parameters \protect\cite{c3}}

\begin{figure}[!t]
\includegraphics[width=8cm]{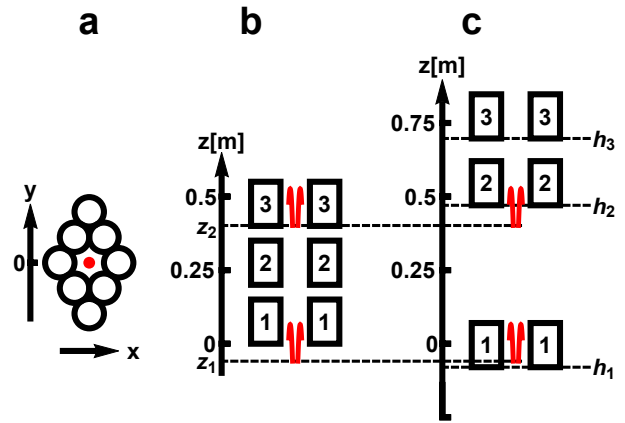}
\caption{The same as figure \protect\ref%
{g1} but for the source mass consisting of 3 parts}
\label{g2}
\end{figure}

In this paper, we propose to divide the source mass not into two halves (as
in the article \cite{c3}), but into three parts. The calculation showed that
even in this case, 24 cylinders are enough for the gradient of the source
mass gravitational field to compensate the gradient of the earth's field. A
symmetrical in the horizontal plane configuration of the source mass is
shown in figure \ref{g2}.

In a $C-$configuration, according to \cite{c12} and unlike \cite{c3} we have
chosen for calculations in this section the distance between the lower and
upper set of cylinders $dh=0.05$m. One could use the local maximum and
minimum of the phase equation (\ref{2.3}) in the coordinate and velocity
space%
\begin{equation}
z_{1}=-0.059\text{m, }z_{2}=0.402\text{m,}v_{z1}=v_{z2}=v=1.563\text{m s}%
^{-1},  \label{19}
\end{equation}%
from which atomic clouds of the 1st and 2nd AI should be launched. The
phases of the interferometers equation (\ref{12}) will be equal\qquad \qquad
\qquad \qquad \qquad \qquad \qquad \qquad \qquad \qquad \qquad 
\begin{equation}
\left\{ \phi _{s}^{\left( C\right) }\left( z_{1},v\right) ,\phi _{s}^{\left(
C\right) }\left( z_{2},v\right) \right\} =\left\{ 0.144901\text{rad}%
,-0.150482\text{rad}\right\} .  \label{20}
\end{equation}%
One should underline that the extremas equation (\ref{19}) are different
from the absolute maxima and minima of the phase $\phi _{s}^{\left( C\right)
}\left( z,v\right) $, which were previously considered for the cuboid source
mass in the references \cite{c5,c10}.

The launching velocity in equation (\ref{19}) is close to the velocity of
the atomic fountain \cite{c24} $gT.$ differing from it only in the third
digit, 
\begin{equation}
\delta v=v-gT\approx -5.8\cdot 10^{-3}\text{m s}^{-1}.  \label{21.1}
\end{equation}%
From equations (\ref{36}), (\ref{S.4a}), and (\ref{S.7a}) one sees that
deviation $\delta v$ could only slightly depend on the interrogation time $%
T. $ The difference equation (\ref{21.1}), however, is sufficient to exclude
the parasitic signal\ \cite{c25}, which occurs when atoms interact with a
Raman pulse having an opposite sign of the effective wave vector. Indeed,
the Raman frequency detuning for the parasitic signal $\delta =2k\delta
v\approx -2\cdot 10^{5}$s$^{-1}.$ If the duration of the $\pi -$pulse $\tau
\thicksim 60\mu $s, then the absolute value of the detuning $\delta $ is an
order of magnitude greater than the inverse pulse duration, and the
probability of excitation of atoms by a parasitic Raman field is negligible,
is estimated to be about 4\%.

\begin{figure}[!t]
\includegraphics[width=16cm]{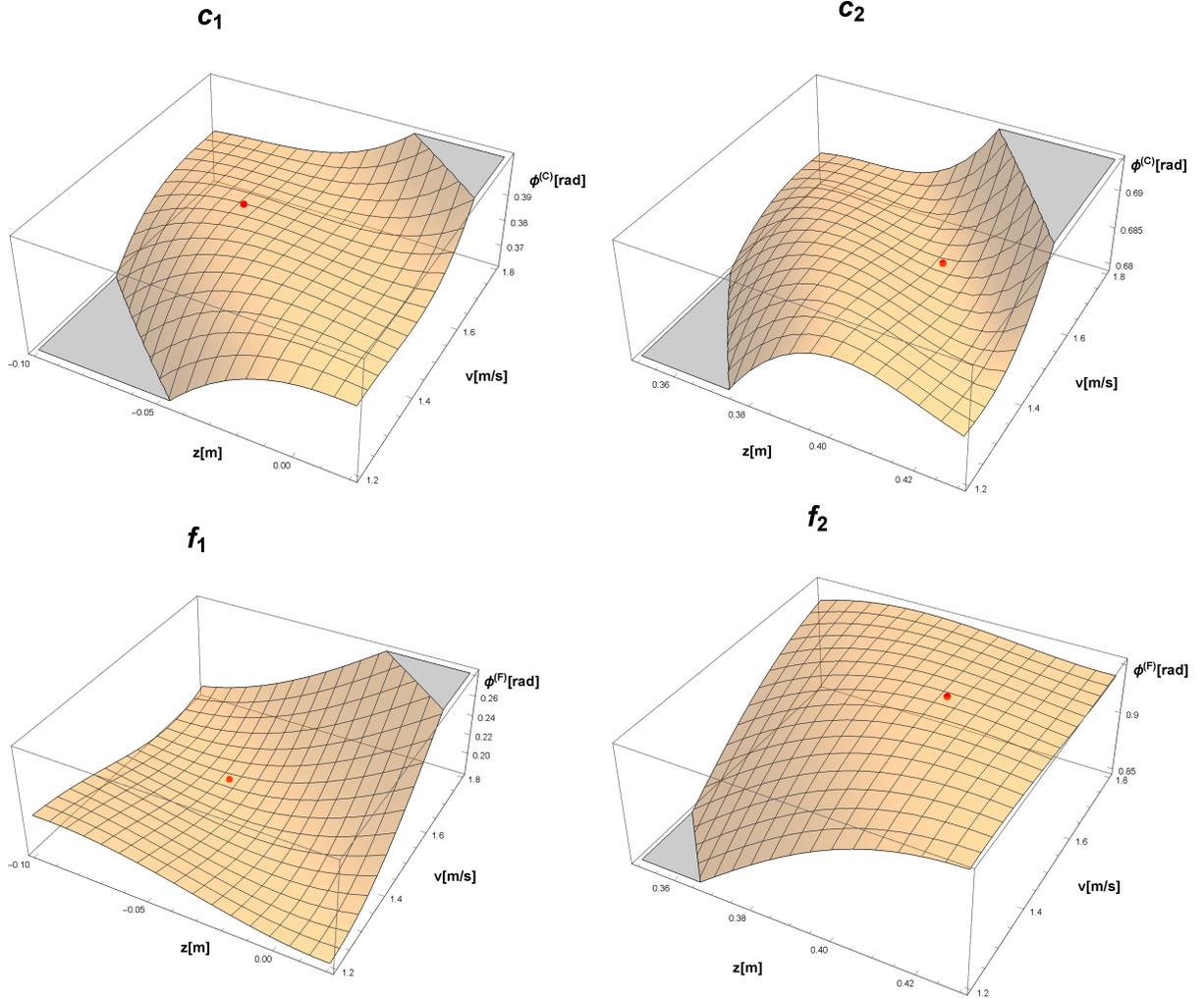}
\caption{Phases of AIS $1$ and $2$ in the
vicinity of points $\left\{ z_{1},v\right\} $ and $\left\{ z_{2},v\right\} $%
, given in (eq. \protect\ref{19}) for $C-$ and $F-$configurations of the
source mass. In the expression for the phase, we kept only the terms that
depend on the vertical components of the atomic coordinates and velocities,
i.e., we put that $\protect\phi ^{\left( C,F\right) }\left( z,v\right)
=k\Gamma _{E33}T^{2}\left[ z+v\left( T+T_{1}\right) \right] +\protect\phi %
_{s}^{\left( C,F\right) }\left( z,v\right) .$ Extremas are shown in red. One
sees that the point $\left\{ z_{1,}v\right\} $ is maximum in $C-$%
configuration and minimum in $F-$configuration and, on the contrary, the
point $\left\{ z_{2},v\right\} $ is the minimum in $C-$configuration and
maximum in $F-$configuration.}
\label{f0}
\end{figure}

Let us now consider the $F-$configuration, see figure \ref{g2}(c) We are
looking for an arrangement of parts of the source mass where the points (\ref%
{19}) are still extreme, i.e. the coordinates of the parts of the source
mass $\left\{ h_{1},h_{2},h_{3}\right\} $ are the roots of a system of 4
equations (equation (\ref{n1})) with a constraint (equation (\ref{n0})).
There can be at least 2 such solutions. We have found and offer it for use a
numerical solution 
\begin{equation}
\left\{ h_{1},h_{2},h_{3}\right\} =\left\{ -0.080\text{m},0.470\text{m},0.697%
\text{m}\right\} ,  \label{21}
\end{equation}%
when the point $\left\{ z_{1},v\right\} $ becomes the local minimum, and the
point $\left\{ z_{2},v\right\} $ becomes the local maximum of the AI phase,%
\begin{equation}
\left\{ \phi _{s}^{\left( F\right) }\left( z_{1},v\right) ,\phi _{s}^{\left(
F\right) }\left( z_{2},v\right) \right\} =\left\{ -0.065580\text{rad,}%
0.107651\text{rad}\right\}  \label{22}
\end{equation}%
Using equations (\ref{1}), (\ref{2}), (\ref{20}), and (\ref{22}) one gets
for PDD 
\begin{equation}
\Delta ^{\left( 2\right) }\phi =0.468614\text{rad.}  \label{23}
\end{equation}%
The dependencies of the AIs phase (equation (\ref{2.4})) near the extremas
are shown in figure \ref{f0}.

\onecolumngrid%

\twocolumngrid%

Details of the calculations of the error budget one can find in Appendix \ref%
{m2}. For SDs equation (\ref{17}) achieved in \cite{c3} using equation (\ref%
{S.2}) one arrives to the RSD and the shift 
\begin{subequations}
\label{24}
\begin{gather}
\sigma \left( \delta \Delta ^{\left( 2\right) }\phi \right) =75\text{ppm}%
\left[ 1+1.05\cdot 10^{\symbol{94}15}\left( \Gamma _{E31}^{2}+\Gamma
_{E32}^{2}\right) \right] ^{1/2}\text{,}  \label{24a} \\
s\left( \delta \Delta ^{\left( 2\right) }\phi \right) =120\text{ppm.}
\label{24b}
\end{gather}%
The non-diagonal elements of the gravity-gradient tensor of the earth field, 
$\Gamma _{E31}$ and $\Gamma _{E32},$ can be neglected with an accuracy of
not more than 10\% if 
\end{subequations}
\begin{equation}
\sqrt{\Gamma _{E31}^{2}+\Gamma _{E32}^{2}}<6.5\text{E}.  \label{25}
\end{equation}

\subsection{\label{s4.2}Using suggested Rossi parameters \protect\cite{c15}}

We have already mentioned above that Rosi proposed and studied \cite{c15} a
new approach to the measurement of $G$ with an accuracy of 10 ppm, based on
the technique of eliminating the gravity-gradient terms \cite{c17}. In
addition to the new technique, estimates have been performed for the source
mass weight increased to the 13 tons, time separation between Raman pulses
increased to 
\begin{equation}
T=243\text{ms,}  \label{34.1}
\end{equation}%
and the uncertainty of the velocity of atomic clouds reduced to 
\begin{subequations}
\label{35}
\begin{eqnarray}
\sigma \left( v_{xjI}\right) &=&\sigma \left( v_{yjI}\right) =2\text{mm s}%
^{-1},  \label{35a} \\
\sigma \left( v_{zjI}\right) &=&0.3\text{mm s}^{-1}\text{.}  \label{35b}
\end{eqnarray}%
In this section, we tested our method of dividing the source mass into 3
parts for parameters close to those proposed in \cite{c15} and for the
source mass consisting of cylinders used in \cite{c3}. We chose the location
of the cylinders on three floors shown in figure \ref{g3}. It is easy to see
that the cylinders are still positioned symmetrically in the horizontal
plane, and their total weight only slightly exceeds 13 tons. This
arrangement of the cylinders is a natural generalization of the geometry
chosen in the \cite{c3}.

\begin{figure}[!t]
\includegraphics[width=8cm]{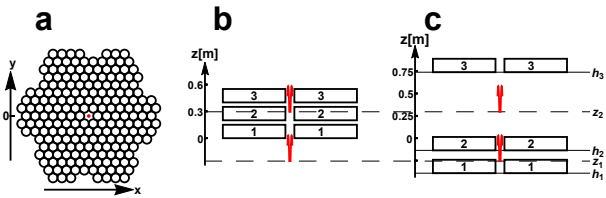}
\caption{The same as figure \protect\ref%
{g2} but for the 13-ton source mass.}
\label{g3}
\end{figure}

In a $C-$configuration, one could use the local maximum and minimum of the
phase (equation (\ref{2.3})) in the coordinate and velocity space 
\end{subequations}
\begin{equation}
z_{1}=-0.257\text{m, }z_{2}=0.296\text{m,}v_{z1}=v_{z2}=v=2.377\text{m s}%
^{-1},  \label{36}
\end{equation}%
from which it is necessary to launch atomic clouds of the first and second
AI. The phases of the interferometers equation (\ref{12}) will be equal%
\begin{equation}
\left\{ \phi _{s}^{\left( C\right) }\left( z_{1},v\right) ,\phi _{s}^{\left(
C\right) }\left( z_{2},v\right) \right\} =\left\{ 1.61119\text{rad},-1.53732%
\text{rad}\right\} .  \label{37}
\end{equation}

Let us now consider the $F-$configuration, see figure \ref{g3}(c). We are
looking for an arrangement of parts of the source mass where the points
(equation (\ref{36})) are still extreme, i.e. the coordinates of the parts
of the source mass $\left\{ h_{1},h_{2},h_{3}\right\} $ are the roots of a
system of 4 equations (equation (\ref{n1})) with a constraint (equation (\ref%
{n0})). There can be at least 2 such solutions. We have found and offer it
for use a numerical solution 
\begin{equation}
\left\{ h_{1},h_{2},h_{3}\right\} =\left\{ -0.377\text{m},-0.153\text{m}%
,0.561\text{m}\right\} ,  \label{38}
\end{equation}%
when the point $\left\{ z_{1},v\right\} $ becomes the local minimum, and the
point $\left\{ z_{2},v\right\} $ becomes the local maximum of the AI phase,%
\begin{equation}
\left\{ \phi _{s}^{\left( F\right) }\left( z_{1},v\right) ,\phi _{s}^{\left(
F\right) }\left( z_{2},v\right) \right\} =\left\{ -0.724326\text{rad,}%
-0.015028\text{rad}\right\}  \label{39}
\end{equation}%
Using equations (\ref{1}), (\ref{2}), (\ref{20}), and (\ref{22}) one gets
for PDD 
\begin{equation}
\Delta ^{\left( 2\right) }\phi =3.85782\text{rad.}  \label{40}
\end{equation}%
Details of the calculations of the error budget one can find in Appendix \ref%
{m3}. For SDs (equations (\ref{17a}) (\ref{17b}), (\ref{35})) using
(equation (\ref{S.3})) one arrives to the RSD and the shift 
\begin{subequations}
\label{41}
\begin{align}
\sigma \left( \delta \Delta ^{\left( 2\right) }\phi \right) & =23\text{ppm}%
\left[ 1+5.84\cdot 10^{14}\left( \Gamma _{E31}^{2}+\Gamma _{E32}^{2}\right) %
\right] ^{1/2}\text{,}  \label{41a} \\
s\left( \delta \Delta ^{\left( 2\right) }\phi \right) & =45\text{ppm.}
\label{41b}
\end{align}%
The non-diagonal elements of the gravity-gradient tensor of the earth field, 
$\Gamma _{E31}$ and $\Gamma _{E32},$ can be neglected with an accuracy of
not more than 10\% if 
\end{subequations}
\begin{equation}
\sqrt{\Gamma _{E31}^{2}+\Gamma _{E32}^{2}}<19\text{E}.  \label{42}
\end{equation}

\section{\label{s5}Conclusion}

This article is devoted to the calculation of the error budget in the
measurement of the Newtonian gravitational constant $G$ by atomic
interferometry methods. Using the technique \cite{c14}, we obtained
expressions for the gravitational field of the cylinder, which is used in
these measurements.

Despite the compensation of the gradient of the earth gravitational field at
the points of apogees of the atomic trajectories achieved in the article 
\cite{c3}, an absence of this compensation along the entire trajectory leads
to the influence of the earth's field on the $G$ measurement accuracy. To
overcome this influence, we propose to use source mass divided on three or
more parts.

The main attention in this article is paid to the calculation of SD and the
shift of the PDD due to the uncertainties of the mean values of the initial
coordinates and the velocities of atomic clouds $\left\{ \delta \mathbf{x}%
,\delta \mathbf{v}\right\} $. We propose to include in the error budget new
terms. They are originated from the quadratic dependence of the variation of
the AI phases on $\left\{ \delta \mathbf{x},\delta \mathbf{v}\right\} $. The
shift arises only after including those terms. At the conditions realized in
the article \cite{c3}, calculations brings us to the shift (equation (\ref%
{18b})) and to the opposite relative correction $\Delta G/G=-199$ppm, which
is larger than corrections considered in \cite{c3}. After including this
correction, the value of the gravitational constant $G$ should be shifted to 
\begin{equation}
G=6.67058\ast 10^{-11}\text{m}^{3}\text{kg}^{-1}\text{s}^{-2}  \label{46.1}
\end{equation}%
from the value $G=6.67191\ast 10^{-11}$m$^{3}$kg$^{-1}$s$^{-2}$ measured in 
\cite{c3}. Monte Carlo simulation was used in \cite{c3} to determine the
constant $G$. In principle, if one includes in the simulation the variations
in the spatial centers of atomic clouds and the centers of the atomic
velocity distribution, the shift (equation (\ref{18b})) would be included in
the averaging over random samples of the atomic coordinates and velocities.
However, these variations, according to \cite{c3}, were not included in the
Monte Carlo simulation, which allows us to suggest shifting the measurement
result of the constant $G$ to the value (equation (\ref{46.1})).

Another discrepancy with article \cite{c3} is the measurement accuracy.
According to our calculation for the atomic coordinates' and velocities' SDs
(equation (\ref{17})), which were achieved in \cite{c3}, the measurement
accuracy of $G$ should not be less than $275$ppm, see equation (\ref{18a}),
while according to \cite{c3} the total accuracy was $148$ppm.

We propose to generalize the method \cite{c12} as follows. In the $C-$%
configuration, when all the parts of the source mass are pieced together, we
look for local extremas of the total phase of the atomic interferometer
equation (\ref{2.3}), using the equations (\ref{2.4}), (\ref{12})-(\ref{12.2}%
). In addition to the atomic coordinates, which were varied in \cite{c12},
we vary also the atomic velocities. Calculations have shown that the atomic
velocities at the points of local maximum and minimum coincide. The task now
is to ensure that these found points remain extreme in the $F-$%
configuration. At the same time, in order for the contribution to the PDD
from the $F-$configuration to be positive, the former maximum point $\left(
z_{1},v_{z_{1}}\right) $ must become the minimum point, and the former
minimum point $\left( z_{2},v_{z_{2}}\right) $ must become the maximum
point. The necessary condition for this is that the phase of the atomic
interferometer, as a function of the positions of the source mass parts,
must satisfy a system of four equations (equation (\ref{n1})) with a
restraint (equation (\ref{n0})), i.e., at least, the phase must be a
function of 3 parameters of the source mass. In contrast to \cite{c12,c3},
we propose to divide it into 3 parts, and to choose for the parameters the 3
vertical coordinates of the source mass parts $\left\{
h_{1},h_{2},h_{3}\right\} $.

We considered this procedure in Section \ref{s4} for the \newpage

\onecolumngrid%

\begin{table}[tbp]
\caption{Error budgets summaries for different numbers of the source mass
parts and different numbers of cylinders. We neglect contribution to the RSD
from the non diagonal matrix elements of the Earth's field gravity-gradient
tensor. Budgets in the columns 3-5,7,8 are built in this article.}
\label{t0}%
\begin{tabular}{|llllllll|}
\hline
Number of source mass parts & $2$ & $2$ & $3$ & $4$ & $1$ & $3$ & $4$ \\ 
Source mass [kg] & $516$ & $516$ & $516$ & $688$ & $13000$ & $13022$ & $%
13065 $ \\ 
Uncertainty of the vertical position $\sigma \left( z_{jI}\right) $ [m] & $%
10^{-4}$ & $10^{-4}$ & $10^{-4}$ & $10^{-4}$ & $10^{-4}$ & $10^{-4}$ & $%
10^{-4}$ \\ 
Uncertainty of the horizontal position $\sigma \left( x_{jI}\right) =\sigma
\left( y_{jI}\right) $ [m] & $10^{-3}$ & $10^{-3}$ & $10^{-3}$ & $10^{-3}$ & 
$10^{-3}$ & $10^{-3}$ & $10^{-3}$ \\ 
Uncertainty of the vertical velocity $\sigma \left( v_{zjI}\right) $ [m/s] & 
$3\cdot 10^{-3}$ & $3\cdot 10^{-3}$ & $3\cdot 10^{-3}$ & $3\cdot 10^{-3}$ & $%
2\cdot 10^{-3}$ & $2\cdot 10^{-3}$ & $2\cdot 10^{-3}$ \\ 
Uncertainty of the horizontal velocity $\sigma \left( v_{xjI}\right) =\sigma
\left( v_{yjI}\right) $ [m/s] & $6\cdot 10^{-3}$ & $6\cdot 10^{-3}$ & $%
6\cdot 10^{-3}$ & $6\cdot 10^{-3}$ & 3$\cdot 10^{-4}$ & 3$\cdot 10^{-4}$ & 3$%
\cdot 10^{-4}$ \\ 
Interrogation time [ms] & $160$ & $160$ & $160$ & $160$ & $243$ & $243$ & $%
243$ \\ 
Source mass configurations & Figure \ref{g1} & Figure \ref{g1} & Fig. \ref%
{g2} & Figure \ref{f1} &  & Figure \ref{g3} & Figure \ref{f2} \\ 
PDD [rad] & $0.547870$ & $0.530535$ & $0.468599$ & $0.523494$ &  & $3.85769$
& $4.72602$ \\ 
RSD [ppm] & $148$ & $275$ & $75$ & $60$ & $10$ & $23$ & $17$ \\ 
Shift [ppm] &  & $199$ & $120$ & $96$ &  & $45$ & $35$ \\ \hline
Reference & \cite{c3} &  &  &  & \cite{c15} &  &  \\ \hline
\end{tabular}%
\end{table}
\twocolumngrid%

same number of cylinders as in \cite{c12,c3}. \textbf{In fact, it is shown
that a simple redistribution of the cylinders between the floors of the
source mass should lead to an improvement in the accuracy of the Newton
gravitational constant measurement by a factor of }$\boldsymbol{3.7}$ 
\textbf{[compare (equations (\ref{18a}) and (\ref{24a}))].}

Following the statement \cite{c15}, that 13-ton source mass can be
implemented in the experiment, we increased the number of cylinders to $606$
(more than $25$ times). At the same time, the PDD increased only $8.4$ times
(compare equations (\ref{23}) and (\ref{40})), and this increase is partly
due to an increase in the delay time between the Raman pulses $T.$ This
example shows that an increase in the weight of the source mass does not
even lead to a proportional signal increase. More promising here is an
increase in the signal due to the larger value of the effective wave vector $%
k,$ longer interrogation time $T,$ and the optimal aspect ratio of the
source mass. Due to these factors we predicted \cite{c5} PDD $\Delta
_{s}^{\left( 2\right) }\phi =386.527$rad even for a source mass $M=1080$kg.

We showed that for the parameters chosen for estimates in \cite{c15}, our
methods of dividing a source mass in 3 parts, leads to the measurement
accuracy $23$ppm, comparable with $10$ppm accuracy predicted in \cite{c15}
for an alternative method \cite{c17} of eliminating the gravity gradient
tensor.

We also performed calculations for the source mass consisting of four
quarters. In this case, in the $F-$configuration, the first and second
floors of the source mass can be located under the first AI, while the third
and fourth floors above the second AI. As a result, the contribution to the
PDD from the point of the minimum $\left( z_{1},v_{z_{1}}\right) $, or from
the point of the maximum $\left( z_{2},v_{z_{2}}\right) $ will increase
leading possibly to the smaller value of RSD. If one still uses 24
cylinders, as in \cite{c12,c3}, after dividing them in four equal quarters,
the gravitational field is too weak to compensate the gradient of the
earth's field. As a result, there are no local extremas in the $F-$%
configuration. To get extremas one needs at least 32 cylinders. It is
reasonable for us to consider this situation, since even larger amount of
cylinders have been used , for example, in the article \cite{c26}. We
arrived to the following result (see Appendix \ref{m4.1} ) 
\begin{subequations}
\label{46.2}
\begin{gather}
\Delta ^{2}\phi =0.523511\text{rad,}  \label{46.2a} \\
\sigma \left( \delta \Delta ^{\left( 2\right) }\phi \right) =60\text{ppm}%
\left[ 1+1.32\cdot 10^{\symbol{94}15}\left( \Gamma _{E31}^{2}+\Gamma
_{E32}^{2}\right) \right] ^{1/2}\text{,}  \label{46.2b} \\
s\left( \delta \Delta ^{\left( 2\right) }\phi \right) =96\text{ppm.}
\label{46.2c}
\end{gather}%
Since we had to change the weight of the source mass, comparison with the
results (equations (\ref{23}) and (\ref{24})) is unfair.

We also considered the case of 4 quarters for 13-ton source mass (see
Appendix \ref{m4.2} ), and arrived to the following result 
\end{subequations}
\begin{subequations}
\label{46.3}
\begin{gather}
\Delta ^{2}\phi =4.72618\text{rad,}  \label{46.3a} \\
\sigma \left( \delta \Delta _{s}^{\left( 2\right) }\phi \right) =17\text{ppm}%
\left[ 1+6.54\cdot 10^{14}\left( \Gamma _{E31}^{2}+\Gamma _{E32}^{2}\right) %
\right] ^{1/2}\text{,}  \label{46.3b} \\
s\left( \delta \Delta _{s}^{\left( 2\right) }\phi \right) =35\text{ppm.}
\label{46.3c}
\end{gather}%
Comparing this with equations (\ref{40}) and (\ref{41}) shows that using a
source mass consisting of 4 quarters could increase PDD in a factor $1.22$
and improve the measurement accuracy in a factor $1.35.$

Results of the our study are summarized in the table \ref{t0}, where the
budgets obtained in references \cite{c3,c15} are also included for
comparison.

Another application of our formulas is the calculation of the systematic
error due to the finite size of atomic clouds and their finite temperature 
\cite{c5,c10}. Let us now assume that $\delta \mathbf{x}$ is the deviation
of the atom from the center of the cloud and $\delta \mathbf{v}$ is the
deviation from the center of the atomic velocity distribution. If the
temperatures are small enough to ignore the Doppler frequency shift, and the
aperture of the optical field is large enough to assume that the areas of
the Raman pulses do not depend on the position of atoms in the cloud, then
the only reason for the dependence of the PDD on $\left\{ \delta \mathbf{x}%
,\delta \mathbf{v}\right\} $ is that the gravitational field $\delta \mathbf{%
g}\left[ \mathbf{x}\left( t\right) \right] $ is not the same for different
atoms in the cloud. In the given $\dfrac{\pi }{2}-\pi -\dfrac{\pi }{2}$ AI,
taking into account the phase dependence on coordinates and velocities,
instead of the usual expression for the probability of excitation of atoms 
\end{subequations}
\begin{equation}
\tilde{P}\left( \phi \right) =\dfrac{1}{2}\left[ 1-\cos \left( \phi \right) %
\right] ,  \label{46.4}
\end{equation}%
one gets the relative population of the upper atomic sublevel averaged over
coordinates and velocities,%
\begin{equation}
P=\dfrac{1}{2}\left\{ 1-\int d\vec{x}d\vec{v}f\left( \vec{x},\vec{v}\right)
\cos \left[ \phi \left( \vec{x},\vec{v}\right) \right] \right\} ,
\label{46.5}
\end{equation}%
where $f\left( \vec{x},\vec{v}\right) $ is the distribution function of
atoms. If the centers of the atomic cloud $\left\{ \vec{x}_{0},\vec{v}%
_{0}\right\} $ are defined as%
\begin{equation}
\left\{ \vec{x}_{0},\vec{v}_{0}\right\} ^{T}=\int d\vec{x}d\vec{v}f\left( 
\vec{x},\vec{v}\right) \left\{ \vec{x},\vec{v}\right\} ^{T},  \label{46.6}
\end{equation}%
then for a sufficiently small radius and a sufficiently low temperature of
the atomic cloud one can use the expansion%
\begin{gather}
\phi \left( \vec{x},\vec{v}\right) =\phi \left( \vec{x}_{0},\vec{v}%
_{0}\right) +\partial _{\vec{x}_{o}}\phi \left( \vec{x}_{0},\vec{v}%
_{0}\right) \delta \vec{x}+\partial _{\vec{v}_{o}}\phi \left( \vec{x}_{0},%
\vec{v}_{0}\right) \delta \vec{v}  \notag \\
+\dfrac{1}{2}\dfrac{\partial ^{2}\phi \left( \vec{x}_{0},\vec{v}_{0}\right) 
}{\partial x_{0m}\partial x_{0n}}\delta x_{m}\delta x_{n}+\dfrac{1}{2}\dfrac{%
\partial ^{2}\phi _{s}\left( \vec{x}_{0},\vec{v}_{0}\right) }{\partial
v_{0m}\partial v_{0n}}\delta v_{m}\delta v_{n}  \notag \\
+\dfrac{\partial ^{2}\phi \left( \vec{x}_{0},\vec{v}_{0}\right) }{\partial
x_{0m}\partial v_{0n}}\delta x_{m}\delta v_{n}.  \label{46.7}
\end{gather}%
Due to the chosen definition of the centers (\ref{46.6}), the linear terms
in the expansion (equation (\ref{46.7})) can be omitted. For statistically
independent distributions on coordinates and velocities, when%
\begin{eqnarray}
\int d\vec{x}d\vec{v}f\left( \vec{x},\vec{v}\right) \delta x_{m}\delta x_{n}
&=&\dfrac{a_{m}^{2}}{2}\delta _{mn},  \label{46.8a} \\
\int d\vec{x}d\vec{v}f\left( \vec{x},\vec{v}\right) \delta v_{m}\delta v_{n}
&=&\dfrac{\bar{v}_{m}^{2}}{2}\delta _{mn},  \label{46.8b} \\
\int d\vec{x}d\vec{v}f\left( \vec{x},\vec{v}\right) \delta x_{m}\delta v_{n}
&=&0,  \label{46.8c}
\end{eqnarray}%
where $a_{m}$ and $\bar{v}_{m}$ are the radius and thermal velocity of the
atomic cloud along the $m-$axis, one gets%
\begin{eqnarray}
P\left( \phi \right) &=&\dfrac{1}{2}\left\{ 1-\cos \left[ \phi \left( \vec{x}%
_{0},\vec{v}_{0}\right) \right] +\dfrac{1}{4}\sin \left[ \phi \left( \vec{x}%
_{0},\vec{v}_{0}\right) \right] \right.  \notag \\
&&\left. \times \left( a_{m}^{2}\dfrac{\partial ^{2}\phi \left( \vec{x}_{0},%
\vec{v}_{0}\right) }{\partial x_{0m}^{2}}+\bar{v}_{m}^{2}\dfrac{\partial
^{2}\phi \left( \vec{x}_{0},\vec{v}_{0}\right) }{\partial v_{0m}^{2}}\right)
\right\} .  \label{46.9}
\end{eqnarray}%
If, despite the non-zero values of the radii and temperatures of atomic
clouds, one wants to use equation (\ref{46.4}) to determine the phase AI $%
\bar{\phi}$, i.e. if $\bar{\phi}$ is the root of the equation $\tilde{P}%
\left( \bar{\phi}\right) =P,$then for $\left\vert \bar{\phi}-\phi \left( 
\vec{x}_{0},\vec{v}_{0}\right) \right\vert \ll \left\vert \bar{\phi}%
\right\vert $ one finds finally%
\begin{equation}
\bar{\phi}\approx \phi \left( \vec{x}_{0},\vec{v}_{0}\right) +\dfrac{1}{4}%
\left( a_{m}^{2}\dfrac{\partial ^{2}\phi }{\partial x_{m}^{2}}+\bar{v}%
_{m}^{2}\dfrac{\partial ^{2}\phi }{\partial v_{m}^{2}}\right) .  \label{47}
\end{equation}

Here we pay attention to the fact that at equal radii, $a_{x}=a_{y}=a_{z},$
and temperatures, $\bar{v}_{x}=\bar{v}_{y}=\bar{v}_{z},$ the systematic
error in (equation (\ref{47})) disappears. This follows from the equations (%
\ref{13c}) and (\ref{13e}) and from the fact that the gravitational field
obeys the Laplace equation and, therefore, the trace of the gravitational
field curvature tensor (equation (\ref{13.1})) $\dsum\limits_{m=1}^{3}\chi
_{s3mm}=0$.

\acknowledgments The author sends appreciations to Dr. M. Prevedeli for the
fruitful discussions and to Dr. G. Rosi for the explanation of some points
in the article \cite{c15}.

\appendix

\section{\label{m1}Source mass consisting of two halves.}

For the source mass and atomic fountains positioning shown in the figure \ref%
{g1}, the table \ref{t1} contains relative contributions to the PDD from two
configurations. Besides the phase values, linear and quadratic terms in the
relative phase variations, due to the uncertainties of atomic coordinates
and velocities, obtained using\ equations (\ref{7}), (\ref{7.1}), and (\ref%
{13}), are also given. Here and below we used the value of the $zz-$%
component of the gravity gradient tensor of the earth field, 
\begin{equation}
\Gamma _{E33}=3.11\cdot 10^{3}\text{E},  \label{S.0}
\end{equation}%
measured in the article \cite{c1.6}. Using data from the table \ref{t1} and
equations (\ref{6a}), (\ref{6b}), (\ref{10}), and (\ref{10.1}), we obtained
following error budget and shift. 
\begin{widetext}

\begin{table}[tbp]
\caption{Source mass consisting of two halves. Relative contributions to the
PDD and error budget for two configurations of
source mass}
\label{t1}%
\begin{tabular}{|c|c|c|}
\hline
Term & C-configuration & F-configuration \\ \hline
$\phi _{s}^{I}\left( z_{i},v_{zi}\right) /\Delta ^{2}\phi $ & $0.354,-0.330$
& $-0.143,0.172$ \\ \hline
Linear in position & $%
\begin{array}{c}
0.322\delta z_{1C}+0.117\delta z_{2C}+7.77\cdot 10^{5} \\ 
\times \left[ \Gamma _{E31}\left( \delta x_{1C}-\delta x_{2C}\right) +\Gamma
_{E32}\left( \delta y_{1C}-\delta y_{2C}\right) \right]%
\end{array}%
$ & $%
\begin{array}{c}
0.132\delta z_{1F}+0.518\delta z_{2F}-7.77\cdot 10^{5} \\ 
\times \left[ \Gamma _{E31}\left( \delta x_{1F}-\delta x_{2F}\right) +\Gamma
_{E32}\left( \delta y_{1F}-\delta y_{2F}\right) \right]%
\end{array}%
$ \\ \hline
Linear in velocity & $%
\begin{array}{c}
0.0377\delta v_{z1C}+0.0150\delta v_{z2C}+1.24\cdot 10^{5} \\ 
\times \left[ \Gamma _{E31}\left( \delta v_{x1C}-\delta v_{x2C}\right)
+\Gamma _{E32}\left( \delta v_{y1C}-\delta v_{y2C}\right) \right]%
\end{array}%
$ & $%
\begin{array}{c}
0.0132\delta v_{z1F}+0.0683\delta v_{z2F}-1.24\cdot 10^{5} \\ 
\times \left[ \Gamma _{E31}\left( \delta v_{x1F}-\delta v_{x2F}\right)
+\Gamma _{E32}\left( \delta v_{y1F}-\delta v_{y2F}\right) \right]%
\end{array}%
$ \\ \hline
Nonlinear in position & $%
\begin{array}{c}
12.3\left( \delta x_{1C}^{2}+\delta y_{1C}^{2}\right) -24.7\delta z_{1C}^{2}
\\ 
+12.2\left( \delta x_{2C}^{2}+\delta y_{2C}^{2}\right) -24.3\delta z_{2C}^{2}%
\end{array}%
$ & $%
\begin{array}{c}
15.5\left( \delta x_{1F}^{2}+\delta y_{1F}^{2}\right) -30.9\delta z_{1F}^{2}
\\ 
+15.7\left( \delta x_{2F}^{2}+\delta y_{2F}^{2}\right) -31.3\delta z_{2F}^{2}%
\end{array}%
$ \\ \hline
Nonlinear in velocity & $%
\begin{array}{c}
0.375\left( \delta v_{x1C}^{2}+\delta v_{y1C}^{2}\right) -0.750\delta
v_{z1C}^{2} \\ 
+0.351\left( \delta v_{x2C}^{2}+\delta v_{y2C}^{2}\right) -0.702\delta
v_{z2C}^{2}%
\end{array}%
$ & $%
\begin{array}{c}
0.451\left( \delta v_{x1F}^{2}+\delta v_{y1F}^{2}\right) -0.901\delta
v_{z1F}^{2} \\ 
+0.468\left( \delta v_{x2F}^{2}+\delta v_{y2F}^{2}\right) -0.937\delta
v_{z2F}^{2}%
\end{array}%
$ \\ \hline
$%
\begin{tabular}{c}
Position-velocity \\ 
cross term%
\end{tabular}%
$ & $%
\begin{array}{c}
3.99\left( \delta v_{x1C}\delta x_{1C}+\delta v_{y1C}\delta y_{1C}\right)
-7.97\delta v_{z1C}\delta z_{1C} \\ 
+4.05\left( \delta v_{x2C}\delta x_{2C}+\delta v_{y2C}\delta y_{2C}\right)
-8.10\delta v_{z2C}\delta z_{2C}%
\end{array}%
$ & $%
\begin{array}{c}
5.10\left( \delta v_{x1F}\delta x_{1F}+\delta v_{y1F}\delta y_{1F}\right)
-10.2\delta v_{z1F}\delta z_{1F} \\ 
+5.08\left( \delta v_{x2F}\delta x_{2F}+\delta v_{y2F}\delta y_{2F}\right)
-10.2\delta v_{z2F}\delta z_{2F}%
\end{array}%
$ \\ \hline
\end{tabular}%
\end{table}

\begin{subequations}
\label{S.1}
\begin{align}
\sigma \left( \Delta _{s}^{\left( 2\right) }\phi \right) & =\left\{
0.104\sigma ^{2}\left( z_{1C}\right) +0.0137\sigma ^{2}\left( z_{2C}\right)
+1.42\cdot 10^{-3}\sigma ^{2}\left( v_{z}{}_{1C}\right) +2.24\cdot
10^{-4}\sigma ^{2}\left( v_{z2C}\right) \right.   \notag \\
& +0.0173\sigma ^{2}\left( z_{1F}\right) +0.269\sigma ^{2}\left(
z_{2F}\right) +1.75\cdot 10^{-4}\sigma ^{2}\left( v_{z1F}\right) +4.66\cdot
10^{-3}\sigma ^{2}\left( v_{z2F}\right)   \notag \\
& +\dsum_{j=1,2}\dsum_{I=C,F}\left[ 6.04\cdot 10^{11}\left( \Gamma
_{E31}^{2}\sigma ^{2}\left( x_{jI}\right) +\Gamma _{E32}^{2}\sigma
^{2}\left( y_{jI}\right) \right) +1.55\cdot 10^{10}\left( \Gamma
_{E31}^{2}\sigma ^{2}\left( v_{xjI}\right) +\Gamma _{E32}^{2}\sigma
^{2}\left( v_{yjI}\right) \right) \right]   \notag \\
& +305\left[ \sigma ^{4}\left( x_{1C}\right) +\sigma ^{4}\left(
y_{1C}\right) \right] +1220\sigma ^{4}\left( z_{1C}\right) +296\left[ \sigma
^{4}\left( x_{2C}\right) +\sigma ^{4}\left( y_{2C}\right) \right]
+1180\sigma ^{4}\left( z_{2C}\right)   \notag \\
& +0.282\left[ \sigma ^{4}\left( v_{x1C}\right) +\sigma ^{4}\left(
v_{y1C}\right) \right] +1.13\sigma ^{4}\left( v_{z1C}\right) +0.247\left[
\sigma ^{4}\left( v_{x2C}\right) +\sigma ^{4}\left( v_{y2C}\right) \right]
+0.987\sigma ^{4}\left( v_{z2C}\right)   \notag \\
& +15.9\left[ \sigma ^{2}\left( x_{1C}\right) \sigma ^{2}\left(
v_{x1C}\right) +\sigma ^{2}\left( y_{1C}\right) \sigma ^{2}\left(
v_{y1C}\right) \right] +63.5\sigma ^{2}\left( z_{1C}\right) \sigma
^{2}\left( v_{z1C}\right)   \notag \\
& +16.4\left[ \sigma ^{2}\left( x_{2C}\right) \sigma ^{2}\left(
v_{x2C}\right) +\sigma ^{2}\left( y_{2C}\right) \sigma ^{2}\left(
v_{y2C}\right) \right] +65.6\sigma ^{2}\left( z_{2C}\right) \sigma
^{2}\left( v_{z2C}\right)   \notag \\
& +478\left[ \sigma ^{4}\left( x_{1F}\right) +\sigma ^{4}\left(
y_{1F}\right) \right] +1910\sigma ^{4}\left( z_{1F}\right) +490\left[ \sigma
^{4}\left( x_{2F}\right) +\sigma ^{4}\left( y_{2F}\right) \right]
+1960\sigma ^{4}\left( z_{2F}\right) +  \notag \\
& +0.406\left[ \sigma ^{4}\left( v_{x1F}\right) +\sigma ^{4}\left(
v_{y1F}\right) \right] +1.63\sigma ^{4}\left( v_{z1F}\right) +0.439\left[
\sigma ^{4}\left( v_{x2F}\right) +\sigma ^{4}\left( v_{y2F}\right) \right]
+1.76\sigma ^{4}\left( v_{z2F}\right)   \notag \\
& +26.0\left[ \sigma ^{2}\left( x_{1F}\right) \sigma ^{2}\left(
v_{x1F}\right) +\sigma ^{2}\left( y_{1F}\right) \sigma ^{2}\left(
v_{y1F}\right) \right] +104\sigma ^{2}\left( z_{1F}\right) \sigma ^{2}\left(
v_{z1F}\right)   \notag \\
& \left. +25.8\left[ \sigma ^{2}\left( x_{2F}\right) \sigma ^{2}\left(
v_{x2F}\right) +\sigma ^{2}\left( y_{2F}\right) \sigma ^{2}\left(
v_{y2F}\right) \right] +103\sigma ^{2}\left( z_{2F}\right) \sigma ^{2}\left(
v_{z2F}\right) \right\} ^{1/2},  \label{S.1a} \\
s\left( \Delta _{s}^{\left( 2\right) }\phi \right) & =12.3\left[ \sigma
^{2}\left( x_{1C}\right) +\sigma ^{2}\left( y_{1C}\right) \right]
-24.7\sigma ^{2}\left( z_{1C}\right) +12.2\left[ \sigma ^{2}\left(
x_{2C}\right) +\sigma ^{2}\left( y_{2C}\right) \right] -24.3\sigma
^{2}\left( v_{z2C}\right)   \notag \\
& +0.375\left[ \sigma ^{2}\left( v_{x1C}\right) +\sigma ^{2}\left(
v_{y1C}\right) \right] -0.750\sigma ^{2}\left( v_{z1C}\right) +0.351\left[
\sigma ^{2}\left( v_{x2C}\right) +\sigma ^{2}\left( v_{y2C}\right) \right]
-0.702\sigma ^{2}\left( v_{z2C}\right)   \notag \\
& +15.5\left[ \sigma ^{2}\left( x_{1F}\right) +\sigma ^{2}\left(
y_{1F}\right) \right] -30.9\sigma ^{2}\left( z_{1F}\right) +15.7\left[
\sigma ^{2}\left( x_{2F}\right) +\sigma ^{2}\left( y_{2F}\right) \right]
-31.3\sigma ^{2}\left( z_{2F}\right)   \notag \\
& +0.451\left[ \sigma ^{2}\left( v_{x1F}\right) +\sigma ^{2}\left(
v_{y1F}\right) \right] -0.902\sigma ^{2}\left( v_{z1F}\right) +0.468\left[
\sigma ^{2}\left( v_{x2F}\right) +\sigma ^{2}\left( v_{y2F}\right) \right]
-0.937\sigma ^{2}\left( v_{z2F}\right) .  \label{S.1b}
\end{align}
\end{subequations}
\end{widetext}

\section{\label{m2}Source mass consisting of three parts.}

For the source mass and atomic fountains positioning shown in the figure \ref%
{g2}, using equation (\ref{S.0}) we arrived to the relative contributions to
the PDD listed in the table \ref{t2}. One sees that despite the choice of
extreme points, linear dependences on $\left\{ \delta z_{jI},\delta
v_{zjI}\right\} $in the phase variation do not completely disappear. This is
because extremas (equation (\ref{19})) and positions of the source mass
parts in the $F-$configuration (equation (\ref{21})) were found
approximately. Here and below negligible linear terms will be excluded from
the calculation. Using data from this table and equations (\ref{6a}), (\ref%
{6b}), (\ref{10}), and (\ref{10.1}), we obtained following error budget and
shift%
\begin{widetext}
\begin{table}[tbp]
\caption{The same as in table \protect\ref{t1} but for the source mass
concicting of three parts}%
\label{t2}%
\begin{tabular}{|c|c|c|}
\hline
Term & C-configuration & F-configuration \\ \hline
$\phi _{s}^{I}\left( z_{i},v_{zi}\right) /\Delta ^{2}\phi $ & $0.309,-0.321$
& $-0.140,0.230$ \\ \hline
Linear in position & $%
\begin{array}{c}
10^{-7}\left( -1.52\delta z_{1C}+1.28\delta z_{2C}\right) +8.80\cdot 10^{5}
\\ 
\times \left[ \Gamma _{E31}\left( \delta x_{1C}-\delta x_{2C}\right) +\Gamma
_{E32}\left( \delta y_{1C}-\delta y_{2C}\right) \right]%
\end{array}%
$ & $%
\begin{array}{c}
10^{-8}\left( 1.29\delta z_{1F}+3.57\delta z_{2F}\right) -8.80\cdot 10^{5}
\\ 
\times \left[ \Gamma _{E31}\left( \delta x_{1F}-\delta x_{2F}\right) +\Gamma
_{E32}\left( \delta y_{1F}-\delta y_{2F}\right) \right]%
\end{array}%
$ \\ \hline
Linear in velocity & $%
\begin{array}{c}
10^{-8}\left( -3.38\delta v_{z1C}+2.18\delta v_{z2C}\right) +1.41\cdot 10^{5}
\\ 
\times \left[ \Gamma _{E31}\left( \delta v_{x1C}-\delta v_{x2C}\right)
+\Gamma _{E32}\left( \delta v_{y1C}-\delta v_{y2C}\right) \right]%
\end{array}%
$ & $%
\begin{array}{c}
10^{-9}\left( -2.97\delta v_{z1F}-5.26\delta v_{z2F}\right) -1.41\cdot 10^{5}
\\ 
\times \left[ \Gamma _{E31}\left( \delta v_{x1F}-\delta v_{x2F}\right)
+\Gamma _{E32}\left( \delta v_{y1F}-\delta v_{y2F}\right) \right]%
\end{array}%
$ \\ \hline
Nonlinear in position & $%
\begin{array}{c}
3.36\delta x_{1C}^{2}+6.22\delta y_{1C}^{2}-9.58\delta z_{1C}^{2} \\ 
+5.00\delta x_{2C}^{2}+8.07\delta y_{2C}^{2}-13.1\delta z_{2C}^{2}%
\end{array}%
$ & $%
\begin{array}{c}
13.3\delta x_{1F}^{2}+16.2\delta y_{1F}^{2}-29.5\delta z_{1F}^{2} \\ 
+6.26\delta x_{2F}^{2}+9.15\delta y_{2F}^{2}-15.4\delta z_{2F}^{2}%
\end{array}%
$ \\ \hline
Nonlinear in velocity & $%
\begin{array}{c}
0.126\delta v_{x1C}^{2}+0.215\delta v_{y1C}^{2}-0.341\delta v_{z1C}^{2} \\ 
+0.117\delta v_{x2C}^{2}+0.203\delta v_{y2C}^{2}-0.320\delta v_{z2C}^{2}%
\end{array}%
$ & $%
\begin{array}{c}
0.362\delta v_{x1F}^{2}+0.443\delta v_{y1F}^{2}-0.805\delta v_{z1F}^{2} \\ 
+0.207\delta v_{x2F}^{2}+0.297\delta v_{y2F}^{2}-0.504319\delta v_{z2F}^{2}%
\end{array}%
$ \\ \hline
$%
\begin{tabular}{c}
Position-velocity \\ 
cross term%
\end{tabular}%
\ $ & $%
\begin{array}{c}
1.07\delta v_{x1C}\delta x_{1C}+1.99\delta v_{y1C}\delta y_{1C}-3.06\delta
v_{z1C}\delta z_{1C} \\ 
+1.60\delta v_{x2C}\delta x_{2C}+2.58\delta v_{y2C}\delta y_{2C}-4.18\delta
v_{z2C}\delta z_{2C}%
\end{array}%
$ & $%
\begin{array}{c}
4.24\delta v_{x1F}\delta x_{1F}+5.19\delta v_{y1F}\delta y_{1F}-9.43\delta
v_{z1F}\delta z_{1F} \\ 
+2.00\delta v_{x2F}\delta x_{2F}+2.93\delta v_{y2F}\delta y_{2F}-4.93\delta
v_{z2F}\delta z_{2F}%
\end{array}%
$ \\ \hline
\end{tabular}%
\end{table}
\begin{subequations}
\label{S.2}
\begin{align}
\sigma \left( \Delta _{s}^{\left( 2\right) }\phi \right) & =\left\{
10^{10}\dsum_{j=1,2}\dsum_{I=C,F}\left[ 77.4\left( \Gamma _{E31}^{2}\sigma
^{2}\left( x_{jI}\right) +\Gamma _{E32}^{2}\sigma ^{2}\left( y_{jI}\right)
\right) +1.98\left( \Gamma _{E31}^{2}\sigma ^{2}\left( v_{xjI}\right)
+\Gamma _{E32}^{2}\sigma ^{2}\left( v_{yjI}\right) \right) \right] \right.  
\notag \\
& +22.5\sigma ^{4}\left( x_{1C}\right) +77.4\sigma ^{4}\left( y_{1C}\right)
+183\sigma ^{4}\left( z_{1C}\right) +49.9\sigma ^{4}\left( x_{2C}\right)
+130\sigma ^{4}\left( y_{2C}\right) +341\sigma ^{4}\left( z_{2C}\right) + 
\notag \\
& +0.0316\sigma ^{4}\left( v_{x1C}\right) +0.0926\sigma ^{4}\left(
v_{y1C}\right) +0.232\sigma ^{4}\left( v_{z1C}\right) +0.0275\sigma
^{4}\left( v_{x2C}\right) +0.0825\sigma ^{4}\left( v_{y2C}\right)
+0.205\sigma ^{4}\left( v_{z2C}\right)   \notag \\
& +1.15\sigma ^{2}\left( x_{1C}\right) \sigma ^{2}\left( v_{x1C}\right)
+3.96\sigma ^{2}\left( y_{1C}\right) \sigma ^{2}\left( v_{y1C}\right)
+9.39\sigma ^{2}\left( z_{1C}\right) \sigma ^{2}\left( v_{z1C}\right)  
\notag \\
& +2.56\sigma ^{2}\left( x_{2C}\right) \sigma ^{2}\left( v_{x2C}\right)
+6.66\sigma ^{2}\left( y_{2C}\right) \sigma ^{2}\left( v_{y2C}\right)
+17.5\sigma ^{2}\left( z_{2C}\right) \sigma ^{2}\left( v_{z2C}\right) + 
\notag \\
& +352\sigma ^{4}\left( x_{1F}\right) +525\sigma ^{4}\left( y_{1F}\right)
+1740\sigma ^{4}\left( z_{1F}\right) +78.3\sigma ^{4}\left( x_{2F}\right)
+1685\sigma ^{4}y_{2F}+475\sigma ^{4}\left( z_{2F}\right)   \notag \\
& +0.262\sigma ^{4}\left( v_{x1F}\right) +0.392^{4}\left( v_{y1F}\right)
+1.30\sigma ^{4}\left( v_{z1F}\right) +0.0858\sigma ^{4}\left(
v_{x2F}\right) +0.177\sigma ^{4}\left( v_{y2F}\right) +0.509\sigma
^{4}\left( v_{z2F}\right)   \notag \\
& +18.0\sigma ^{2}\left( x_{1F}\right) \sigma ^{2}\left( v_{x1F}\right)
+26.9\sigma ^{2}\left( y_{1F}\right) \sigma ^{2}\left( v_{y1F}\right)
+88.9\sigma ^{2}\left( z_{1F}\right) \sigma ^{2}\left( v_{z1F}\right)  
\notag \\
& \left. +4.01\sigma ^{2}\left( x_{2F}\right) \sigma ^{2}\left(
v_{x2F}\right) +8.58\sigma ^{2}\left( y_{2F}\right) \sigma ^{2}\left(
v_{y2F}\right) +24.3\sigma ^{2}\left( z_{2F}\right) \sigma ^{2}\left(
v_{z2F}\right) \right\} ^{1/2}  \label{S.2a} \\
s\left( \Delta _{s}^{\left( 2\right) }\phi \right) & =3.36\sigma ^{2}\left(
x_{1C}\right) +6.22\sigma ^{2}\left( y_{1C}\right) -9.58\sigma ^{2}\left(
z_{1C}\right) +5.00\sigma ^{2}\left( x_{2C}\right) +8.07\sigma ^{2}\left(
y_{2C}\right) -13.1\sigma ^{2}\left( z_{2C}\right)   \notag \\
& +0.126\sigma ^{2}\left( v_{x1C}\right) +0.215\sigma ^{2}\left(
v_{y1C}\right) -0.341\sigma ^{2}\left( v_{z1C}\right) +0.117\sigma
^{2}\left( v_{x2C}\right) +0.203\sigma ^{2}\left( v_{y2C}\right)
-0.320\sigma ^{2}\left( v_{z2C}\right)   \notag \\
& +13.3\sigma ^{2}\left( x_{1F}\right) +16.2\sigma ^{2}\left( y_{1F}\right)
-29.5\sigma ^{2}\left( z_{1F}\right) +6.26\sigma ^{2}\left( x_{2F}\right)
+9.15\sigma ^{2}\left( y_{2F}\right) -15.4\sigma ^{2}\left( z_{2F}\right)  
\notag \\
& +0.362\sigma ^{2}\left( v_{x1F}\right) +0.443\sigma ^{2}\left(
v_{y1F}\right) -0.805\sigma ^{2}\left( v_{z1F}\right) +0.207\sigma
^{2}\left( v_{x2F}\right) +0.297\sigma ^{2}\left( v_{y2F}\right)
-0.504\sigma ^{2}\left( v_{z2F}\right) .  \label{S.2b}
\end{align}
\end{subequations}
\end{widetext}

\section{\label{m3}13-ton source mass.}

For the source mass and atomic fountains positioning shown in the figure \ref%
{g3}, using equation (\ref{S.0}) we arrived to the relative contributions to
the PDD listed in the table \ref{t3}. Using data from this table and
equations (\ref{6a}), (\ref{6b}), (\ref{10}), and (\ref{10.1}), we obtained
following error budget and shift%
\begin{widetext}
\begin{table}[tbp]
\caption{The same as in table \protect\ref{t1} but for the 13 tons source
mass.}%
\label{t3}%
\begin{tabular}{|c|c|c|}
\hline
Term & C-configuration & F-configuration \\ \hline
$\phi _{s}^{I}\left( z_{i},v_{zi}\right) /\Delta ^{2}\phi $ & $0.418,-0.398$
& $-0.188,-3.90\cdot 10^{-3}$ \\ \hline
Linear in position & $%
\begin{array}{c}
10^{-10}\left( -11.0\delta z_{1C}+8.63\delta z_{2C}\right) +2.47\cdot 10^{5}
\\ 
\times \left[ \Gamma _{E31}\left( \delta x_{1C}-\delta x_{2C}\right) +\Gamma
_{E32}\left( \delta y_{1C}-\delta y_{2C}\right) \right]%
\end{array}%
$ & $%
\begin{array}{c}
10^{-9}\left( -7.36\delta z_{1F}-10.8\delta z_{2F}\right) -2.47\cdot 10^{5}
\\ 
\times \left[ \Gamma _{E31}\left( \delta x_{1F}-\delta x_{2F}\right) +\Gamma
_{E32}\left( \delta y_{1F}-\delta y_{2F}\right) \right]%
\end{array}%
$ \\ \hline
Linear in velocity & $%
\begin{array}{c}
10^{-10}\left( -2.97\delta v_{z1C}+2.09\delta v_{z2C}\right) +5.99\cdot
10^{4} \\ 
\times \left[ \Gamma _{E31}\left( \delta v_{x1C}-\delta v_{x2C}\right)
+\Gamma _{E32}\left( \delta v_{y1C}-\delta v_{y2C}\right) \right]%
\end{array}%
$ & $%
\begin{array}{c}
10^{-9}\left( -1.79\delta v_{z1F}-2.25\delta v_{z2F}\right) -5.99\cdot 10^{4}
\\ 
\times \left[ \Gamma _{E31}\left( \delta v_{x1F}-\delta v_{x2F}\right)
+\Gamma _{E32}\left( \delta v_{y1F}-\delta v_{y2F}\right) \right]%
\end{array}%
$ \\ \hline
Nonlinear in position & $%
\begin{array}{c}
4.61\left( \delta x_{1C}^{2}+\delta y_{1C}^{2}\right) -9.22\delta z_{1C}^{2}
\\ 
+4.29\delta x_{2C}^{2}+4.30\delta y_{2C}^{2}-8.59\delta z_{2C}^{2}%
\end{array}%
$ & $%
\begin{array}{c}
4.24\left[ \delta x_{1F}^{2}+\delta y_{1F}^{2}\right] -8.47\delta z_{1F}^{2}
\\ 
+5.10\left[ \delta x_{2F}^{2}+\delta y_{2F}^{2}\right] -10.2\delta z_{2F}^{2}%
\end{array}%
$ \\ \hline
Nonlinear in velocity & $%
\begin{array}{c}
0.297\left( \delta v_{x1C}^{2}+\delta v_{y1C}^{2}\right) -0.594\delta
v_{z1C}^{2} \\ 
+0.270\left( \delta v_{x2C}^{2}+\delta v_{y2C}^{2}\right) -0.540\delta
v_{z2C}^{2}%
\end{array}%
$ & $%
\begin{array}{c}
0.264\left[ \delta v_{x1F}^{2}+\delta v_{y1F}^{2}\right] -0.527\delta
v_{z1F}^{2} \\ 
+0.326\left[ \delta v_{x2F}^{2}+\delta v_{y2F}^{2}\right] -0.652\delta
v_{z2F}^{2}%
\end{array}%
$ \\ \hline
$%
\begin{tabular}{c}
Position-velocity \\ 
cross term%
\end{tabular}%
\ $ & $%
\begin{array}{c}
2.24\left( \delta v_{x1C}\delta x_{1C}+\delta v_{y1C}\delta y_{1C}\right)
-4.48\delta v_{z1C}\delta z_{1C} \\ 
+2.09\left( \delta v_{x2C}\delta x_{2C}+\delta v_{y2C}\delta y_{2C}\right)
-4.18\delta v_{z2C}\delta z_{2C}%
\end{array}%
$ & $%
\begin{array}{c}
2.06\left[ \delta v_{x1F}\delta x_{1F}+\delta v_{y1F}\delta y_{1F}\right]
-4.12\delta v_{z1F}\delta z_{1F} \\ 
+2.48\left[ \delta v_{x2F}\delta x_{2F}+\delta v_{y2F}\delta y_{2F}\right]
-4.96\delta v_{z2F}\delta z_{2F}%
\end{array}%
$ \\ \hline
\end{tabular}%
\end{table}
\begin{subequations}
\label{S.3}
\begin{align}
\sigma \left( \Delta _{s}^{\left( 2\right) }\phi \right) & =\left\{
10^{9}\dsum_{j=1,2}\dsum_{I=C,F}\left[ 60.8\left( \Gamma _{E31}^{2}\sigma
^{2}\left( x_{jI}\right) +\Gamma _{E32}^{2}\sigma ^{2}\left( y_{jI}\right)
\right) +3.59\left( \Gamma _{E31}^{2}\sigma ^{2}\left( v_{xjI}\right)
+\Gamma _{E32}^{2}\sigma ^{2}\left( v_{yjI}\right) \right) \right] \right.  
\notag \\
& +42.4\sigma ^{4}\left( x_{1C}\right) +42.6\sigma ^{4}\left( y_{1C}\right)
+170\sigma ^{4}\left( z_{1C}\right) +36.9\sigma ^{4}\left( x_{2C}\right)
+37.0\sigma ^{4}\left( y_{2C}\right) +148\sigma ^{4}\left( z_{2C}\right) + 
\notag \\
& +0.176\sigma ^{4}\left( v_{x1C}\right) +0.177\sigma ^{4}\left(
v_{y1C}\right) +0.706\sigma ^{4}\left( v_{z1C}\right) +0.146\left[ \sigma
^{4}\left( v_{x2C}\right) +\sigma ^{4}\left( v_{y2C}\right) \right]
+0.583\sigma ^{4}\left( v_{z2C}\right)   \notag \\
& +5.01\sigma ^{2}\left( x_{1C}\right) \sigma ^{2}\left( v_{x1C}\right)
+5.03\sigma ^{2}\left( y_{1C}\right) \sigma ^{2}\left( v_{y1C}\right)
+20.1\sigma ^{2}\left( z_{1C}\right) \sigma ^{2}\left( v_{z1C}\right)  
\notag \\
& +4.36\sigma ^{2}\left( x_{2C}\right) \sigma ^{2}\left( v_{x2C}\right)
+4.37\sigma ^{2}\left( y_{2C}\right) \sigma ^{2}\left( v_{y2C}\right)
+17.4\sigma ^{2}\left( z_{2C}\right) \sigma ^{2}\left( v_{z2C}\right) + 
\notag \\
& +35.9\left[ \sigma ^{4}\left( x_{1F}\right) +\sigma ^{4}\left(
y_{1F}\right) \right] +144\sigma ^{4}\left( z_{1F}\right) +52.1\sigma
^{4}\left( x_{2F}\right) +52.0\sigma ^{4}y_{2F}+208\sigma ^{4}\left(
z_{2F}\right)   \notag \\
& +0.139\left[ \sigma ^{4}\left( v_{x1F}\right) +\sigma ^{4}\left(
v_{y1F}\right) \right] +0.556\sigma ^{4}\left( v_{z1F}\right) +0.213\sigma
^{4}\left( v_{x2F}\right) +0.212\sigma ^{4}\left( v_{y2F}\right)
+0.850\sigma ^{4}\left( v_{z2F}\right)   \notag \\
& +4.24\left[ \sigma ^{2}\left( x_{1F}\right) \sigma ^{2}\left(
v_{x1F}\right) +\sigma ^{2}\left( y_{1F}\right) \sigma ^{2}\left(
v_{y1F}\right) \right] +17.0\sigma ^{2}\left( z_{1F}\right) \sigma
^{2}\left( v_{z1F}\right)   \notag \\
& \left. +6.15\left[ \sigma ^{2}\left( x_{2F}\right) \sigma ^{2}\left(
v_{x2F}\right) +\sigma ^{2}\left( y_{2F}\right) \sigma ^{2}\left(
v_{y2F}\right) \right] +24.6\sigma ^{2}\left( z_{2F}\right) \sigma
^{2}\left( v_{z2F}\right) \right\} ^{1/2},  \label{S.3a} \\
s\left( \Delta _{s}^{\left( 2\right) }\phi \right) & =4.61\left[ \sigma
^{2}\left( x_{1C}\right) +\sigma ^{2}\left( y_{1C}\right) \right]
-9.22\sigma ^{2}\left( z_{1C}\right) +4.29\sigma ^{2}\left( x_{2C}\right)
+4.30\sigma ^{2}\left( y_{2C}\right) -8.59\sigma ^{2}\left( z_{2C}\right)  
\notag \\
& +0.297\left[ \sigma ^{2}\left( v_{x1C}\right) +\sigma ^{2}\left(
v_{y1C}\right) \right] -0.594\sigma ^{2}\left( v_{z1C}\right) +0.270\left[
\sigma ^{2}\left( v_{x2C}\right) +\sigma ^{2}\left( v_{y2C}\right) \right]
-0.540\sigma ^{2}\left( v_{z2C}\right)   \notag \\
& +4.24\left[ \sigma ^{2}\left( x_{1F}\right) +\sigma ^{2}\left(
y_{1F}\right) \right] -8.47\sigma ^{2}\left( z_{1F}\right) +5.10\left[
\sigma ^{2}\left( x_{2F}\right) +\sigma ^{2}\left( y_{2F}\right) \right]
-10.2\sigma ^{2}\left( z_{2F}\right)   \notag \\
& +0.264\left[ \sigma ^{2}\left( v_{x1F}\right) +\sigma ^{2}\left(
v_{y1F}\right) \right] -0.527\sigma ^{2}\left( v_{z1F}\right) +0.326\left[
\sigma ^{2}\left( v_{x2F}\right) +\sigma ^{2}\left( v_{y2F}\right) \right]
-0.652\sigma ^{2}\left( v_{z2F}\right) .  \label{S.3b}
\end{align}
\end{subequations}
\end{widetext}

\section{Source mass consisting of four quarters}

\subsection{\label{m4.1}Minimal source mass}

We performed calculations for the source mass geometry shown in figure \ref%
{f1}.

\begin{figure}[!t]
\includegraphics[width=8cm]{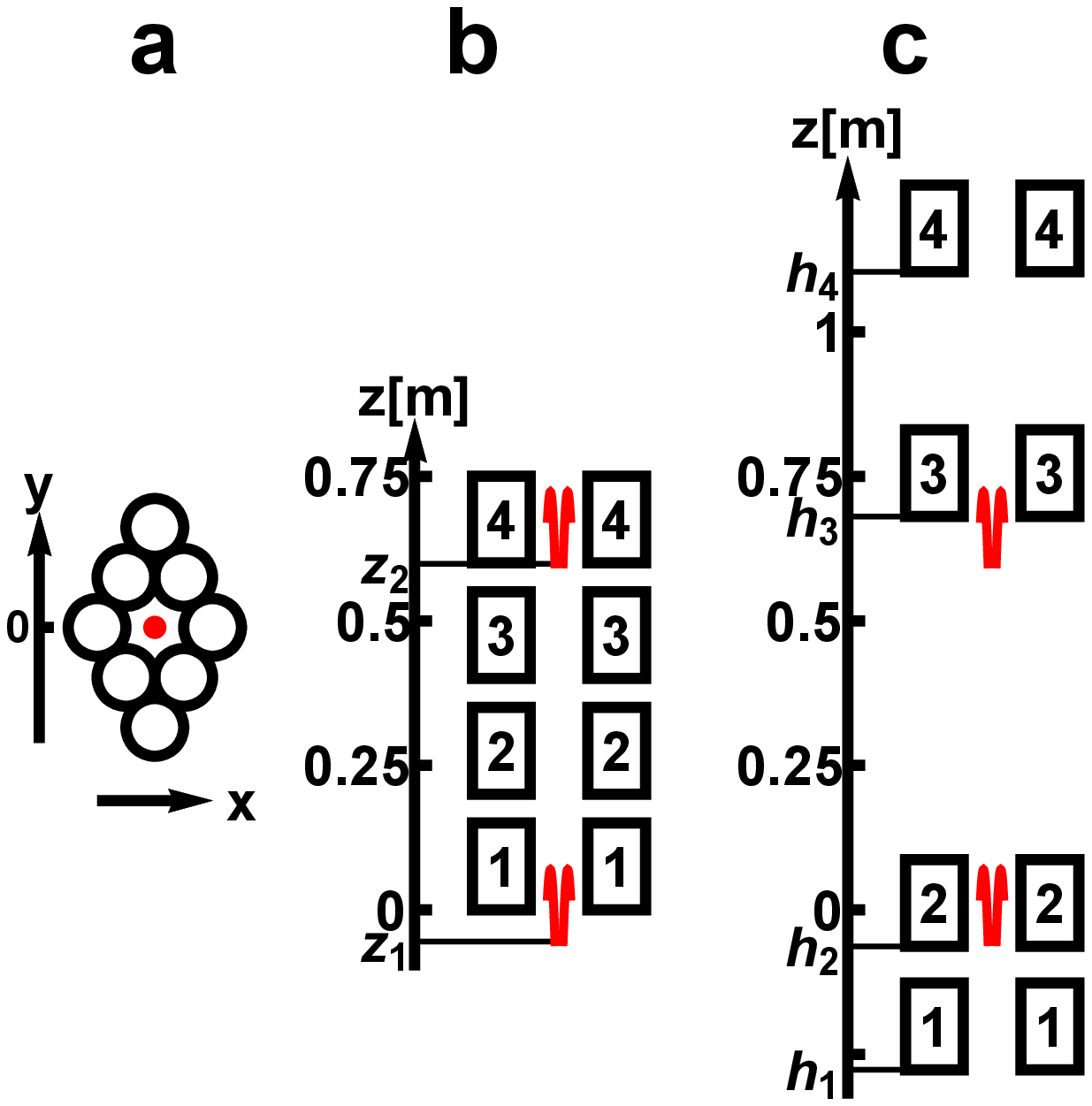}
\caption{The same as figure \protect\ref%
{g2} but for the source mass consisting of 4 quarters.}
\label{f1}
\end{figure}

This is a minimal amount of cylinders used in \cite{c3}, when 4 quarters of
the source mass produce a sufficiently strong gravitational field to make
all atomic variables extreme in the both configurations. Instead of
equations. (\ref{19}), (\ref{20}), (\ref{21})-(\ref{23}) we arrived to the
following results

\begin{subequations}
\begin{gather}
z_{1}=-0.0550\text{m, }z_{2}=0.599\text{m,}v_{z1}=v_{z2}=v=1.563\text{m s$%
^{-1}$},  \label{S.4a} \\
\left\{ \phi _{s}^{\left( C\right) }\left( z_{1},v\right) ,\phi _{s}^{\left(
C\right) }\left( z_{2},v\right) \right\} =\left\{ 0.150759\text{rad}%
,-0.157911\text{rad}\right\} ,  \label{S.4b} \\
\left\{ h_{1},h_{2},h_{3},h_{4}\right\} =\left\{ -0.277\text{m},-0.0632\text{%
m},0.681\text{m,}1.10\text{m}\right\} ,  \label{S.4c} \\
\left\{ \phi _{s}^{\left( F\right) }\left( z_{1},v\right) ,\phi _{s}^{\left(
F\right) }\left( z_{2},v\right) \right\} =\left\{ -0.118580\text{rad, }%
0.096261\text{rad}\right\} ,  \label{S.4d} \\
\Delta ^{2}\phi =0.523511\text{rad.}  \label{S.4e}
\end{gather}%
Relative contributions to the PDD listed now in the table \ref{t4}. Using
data from this table and equations (\ref{6a}), (\ref{6b}), (\ref{10}),and (%
\ref{10.1}), we obtained following error budget and shift 
\begin{widetext}
\begin{table}[tbp]
\caption{The same as in table \protect\ref{t1} but for the source mass
geometry shown in figure \protect\ref{f1}.}%
\label{t4}%
\begin{tabular}{|c|c|c|}
\hline
Term & C-configuration & F-configuration \\ \hline
$\phi _{s}^{I}\left( z_{i},v_{zi}\right) /\Delta ^{2}\phi $ & $0.288,~-0.302$
& -$0.227,~0.184$ \\ \hline
Linear in position & $%
\begin{array}{c}
10^{-8}\left( 3.79\delta z_{1C}-139.\delta z_{2C}\right) +7.88\cdot 10^{5}.
\\ 
\times \left[ \Gamma _{E31}\left( \delta x_{1C}-\delta x_{2C}\right) +\Gamma
_{E32}\left( \delta y_{1C}-\delta y_{2C}\right) \right]%
\end{array}%
$ & $%
\begin{array}{c}
10^{-7}\left( -2.45\delta z_{1F}-27.3\delta z_{2F}\right) -7.88\cdot 10^{5}
\\ 
\times \left[ \Gamma _{E31}\left( \delta x_{1F}-\delta x_{2F}\right) +\Gamma
_{E32}\left( \delta y_{1F}-\delta y_{2F}\right) \right]%
\end{array}%
$ \\ \hline
Linear in velocity & $%
\begin{array}{c}
10^{-9}\left( 6.92\delta v_{z1C}-296\delta v_{z2C}\right) +1.26.\cdot 10^{5}
\\ 
\times \left[ \Gamma _{E31}\left( \delta v_{x1C}-\delta v_{x2C}\right)
+\Gamma _{E32}\left( \delta v_{y1C}-\delta v_{y2C}\right) \right]%
\end{array}%
$ & $%
\begin{array}{c}
10^{-8}\left( -3.91\delta v_{z1F}-4.52\delta v_{z2F}\right) -1.26.\cdot
10^{5} \\ 
\times \left[ \Gamma _{E31}\left( \delta v_{x1F}-\delta v_{x2F}\right)
+\Gamma _{E32}\left( \delta v_{y1F}-\delta v_{y2F}\right) \right]%
\end{array}%
$ \\ \hline
Nonlinear in position & $%
\begin{array}{c}
2.14\delta x_{1C}^{2}+4.53\delta y_{1C}^{2}-6.67\delta z_{1C}^{2} \\ 
+3.48\delta x_{2C}^{2}+6.12\delta y_{2C}^{2}-9.60\delta z_{2C}^{2}%
\end{array}%
$ & $%
\begin{array}{c}
7.03\delta x_{1F}^{2}+9.83\delta y_{1F}^{2}-16.9\delta z_{1F}^{2} \\ 
+9.17\delta x_{2F}^{2}+11.9\delta y_{2F}^{2}-21.0\delta z_{2F}^{2}%
\end{array}%
$ \\ \hline
Nonlinear in velocity & $%
\begin{array}{c}
0.0879\delta v_{x1C}^{2}+0.163\delta v_{y1C}^{2}-0.251\delta v_{z1C}^{2} \\ 
+0.0762\delta v_{x2C}^{2}+0.150\delta v_{y2C}^{2}-0.226\delta v_{z2C}^{2}%
\end{array}%
$ & $%
\begin{array}{c}
0.179\delta v_{x1F}^{2}+0.257\delta v_{y1F}^{2}-0.437\delta v_{z1F}^{2} \\ 
+0.281\delta v_{x2F}^{2}+0.364\delta v_{y2F}^{2}-0.644\delta v_{z2F}^{2}%
\end{array}%
$ \\ \hline
$%
\begin{tabular}{c}
Position-velocity \\ 
cross term%
\end{tabular}%
\ $ & $%
\begin{array}{c}
0.685\delta v_{x1C}\delta x_{1C}+1.45\delta v_{y1C}\delta y_{1C}-2.14\delta
v_{z1C}\delta z_{1C} \\ 
+1.11\delta v_{x2C}\delta x_{2C}+1.96\delta v_{y2C}\delta y_{2C}-3.07\delta
v_{z2C}\delta z_{2C}%
\end{array}%
$ & $%
\begin{array}{c}
2.25\delta v_{x1F}\delta x_{1F}+3.15\delta v_{y1F}\delta y_{1F}-5.40\delta
v_{z1F}\delta z_{1F} \\ 
+2.93\delta v_{x2F}\delta x_{2F}+3.80\delta v_{y2F}\delta y_{2F}-6.73\delta
v_{z2F}\delta z_{2F}%
\end{array}%
$ \\ \hline
\end{tabular}%
\end{table}
\begin{subequations}
\label{S.5}
\begin{align}
\sigma \left( \Delta _{s}^{\left( 2\right) }\phi \right) & =\left\{
10^{10}\dsum_{j=1,2}\dsum_{I=C,F}\left[ 62.0\left( \Gamma _{E31}^{2}\sigma
^{2}\left( x_{jI}\right) +\Gamma _{E32}^{2}\sigma ^{2}\left( y_{jI}\right)
\right) +1.59\left( \Gamma _{E31}^{2}\sigma ^{2}\left( v_{xjI}\right)
+\Gamma _{E32}^{2}\sigma ^{2}\left( v_{yjI}\right) \right) \right] \right.  
\notag \\
& +9.16\sigma ^{4}\left( x_{1C}\right) +41.1\sigma ^{4}\left( y_{1C}\right)
+89.1\sigma ^{4}\left( z_{1C}\right) +24.2\sigma ^{4}\left( x_{2C}\right)
+74.8\sigma ^{4}y_{2C}+184\sigma ^{4}\left( z_{2C}\right)   \notag \\
& +0.0155\sigma ^{4}\left( v_{x1C}\right) +0.0534\sigma ^{4}\left(
v_{y1C}\right) +0.126\sigma ^{4}\left( v_{z1C}\right) +0.0116\sigma
^{4}\left( v_{x2C}\right) +0.0449\sigma ^{4}\left( v_{y2C}\right)
+0.102\sigma ^{4}\left( v_{z2C}\right)   \notag \\
& +0.469\sigma ^{2}\left( x_{1C}\right) \sigma ^{2}\left( v_{x1C}\right)
+2.11\sigma ^{2}\left( y_{1C}\right) \sigma ^{2}\left( v_{y1C}\right)
+4.56\sigma ^{2}\left( z_{1C}\right) \sigma ^{2}\left( v_{z1C}\right)  
\notag \\
& +1.24\sigma ^{2}\left( x_{2C}\right) \sigma ^{2}\left( v_{x2C}\right)
+3.83\sigma ^{2}\left( y_{2C}\right) \sigma ^{2}\left( v_{y2C}\right)
+9.43\sigma ^{2}\left( z_{2C}\right) \sigma ^{2}\left( v_{z2C}\right)  
\notag \\
& +98.9\sigma ^{4}\left( x_{1F}\right) +193\sigma ^{4}\left( y_{1F}\right)
+569\sigma ^{4}\left( z_{1F}\right) +168\sigma ^{4}\left( x_{2F}\right)
+282\sigma ^{4}\left( y_{2F}\right) +885\sigma ^{4}\left( z_{2F}\right) + 
\notag \\
& +0.0643\sigma ^{4}\left( v_{x1F}\right) +0.132\sigma ^{4}\left(
v_{y1F}\right) +0.381\sigma ^{4}\left( v_{z1F}\right) +0.157\sigma
^{4}\left( v_{x2F}\right) +0.265\sigma ^{4}\left( v_{y2F}\right)
+0.830\sigma ^{4}\left( v_{z2F}\right)   \notag \\
& +5.06\sigma ^{2}\left( x_{1F}\right) \sigma ^{2}\left( v_{x1F}\right)
+9.90\sigma ^{2}\left( y_{1F}\right) \sigma ^{2}\left( v_{y1F}\right)
+29.1\sigma ^{2}\left( z_{1F}\right) \sigma ^{2}\left( v_{z1F}\right)  
\notag \\
& \left. +8.60\sigma ^{2}\left( x_{2F}\right) \sigma ^{2}\left(
v_{x2F}\right) +14.4\sigma ^{2}\left( y_{2F}\right) \sigma ^{2}\left(
v_{y2F}\right) +45.3\sigma ^{2}\left( z_{2F}\right) \sigma ^{2}\left(
v_{z2F}\right) \right\} ^{1/2},  \label{S.5a} \\
s\left( \Delta _{s}^{\left( 2\right) }\phi \right) & =2.14\sigma ^{2}\left(
x_{1C}\right) +4.53\sigma ^{2}\left( y_{1C}\right) -6.67\sigma ^{2}\left(
z_{1C}\right) +3.48\sigma ^{2}\left( x_{2C}\right) +6.12\sigma ^{2}\left(
y_{2C}\right) -9.60\sigma ^{2}\left( z_{2C}\right)   \notag \\
& +0.0879\sigma ^{2}\left( v_{x1C}\right) +0.163\sigma ^{2}\left(
v_{y1C}\right) -0.251\sigma ^{2}\left( v_{z1C}\right) +0.0762\sigma
^{2}\left( v_{x2C}\right) +0.150\sigma ^{2}\left( v_{y2C}\right)
-0.226\sigma ^{2}\left( v_{z2C}\right)   \notag \\
& +7.03\sigma ^{2}\left( x_{1F}\right) +9.83\sigma ^{2}\left( y_{1F}\right)
-16.9\sigma ^{2}\left( z_{1F}\right) +9.17\sigma ^{2}\left( x_{2F}\right)
+11.9\sigma ^{2}\left( y_{2F}\right) -21.0\sigma ^{2}\left( z_{2F}\right)  
\notag \\
& +0.179\sigma ^{2}\left( v_{x1F}\right) +0.257\sigma ^{2}\left(
v_{y1F}\right) -0.437\sigma ^{2}\left( v_{z1F}\right) +0.281\sigma
^{2}\left( v_{x2F}\right) +0.364\sigma ^{2}\left( v_{y2F}\right)
-0.644\sigma ^{2}\left( v_{z2F}\right) .  \label{S.5b}
\end{align}
\end{subequations}
\end{widetext}

Substituting here uncertainties of the atomic variables (equation (\ref{17}%
)) one gets instead of (equations (\ref{24}) and (\ref{25})) 
\end{subequations}
\begin{subequations}
\label{S.6}
\begin{gather}
\sigma \left( \delta \Delta ^{\left( 2\right) }\phi \right) =60\text{ppm}%
\left[ 1+1.32\cdot 10^{15}\left( \Gamma _{E31}^{2}+\Gamma _{E32}^{2}\right) %
\right] ^{1/2}\text{,}  \label{S.6a} \\
s\left( \delta \Delta ^{\left( 2\right) }\phi \right) =96\text{ppm,}
\label{S.6b} \\
\sqrt{\Gamma _{E31}^{2}+\Gamma _{E32}^{2}}<6\text{E}.  \label{S.6c}
\end{gather}

\subsection{\label{m4.2}13-ton source mass}

A symmetric in the horizontal plane source mass, which can be divided in 4
quarters and has a minimal total weight exceeding 13 tons, is shown in the
figure \ref{f2}

\begin{figure}[!t]
\includegraphics[width=8cm]{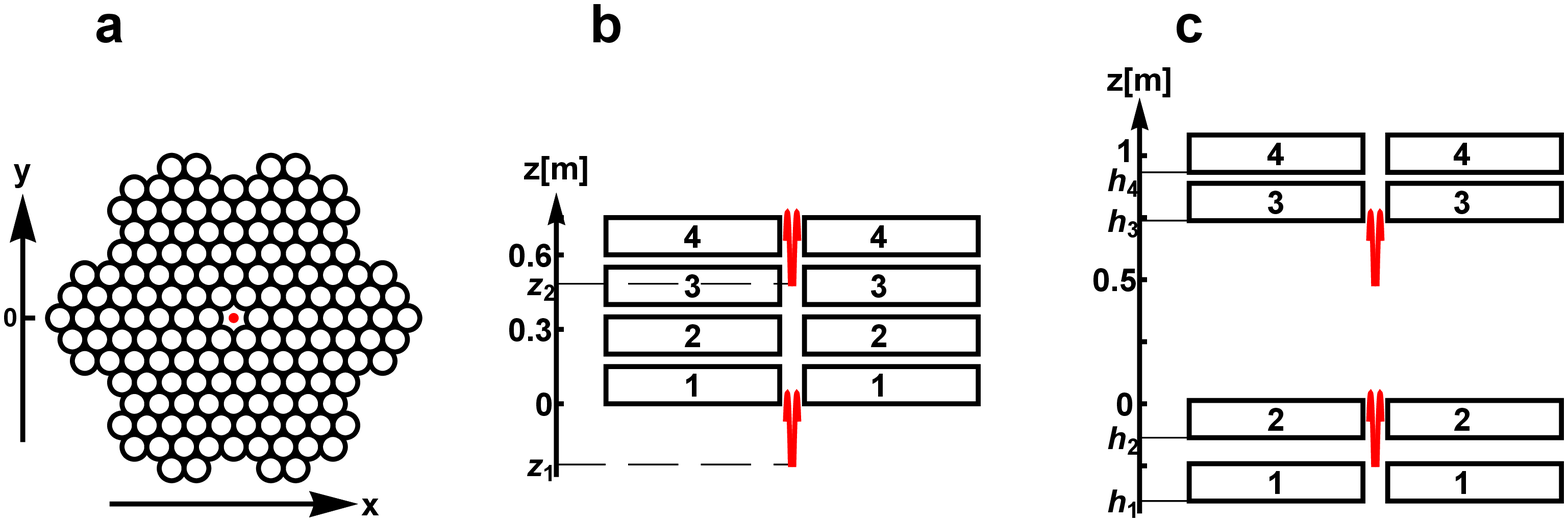}
\caption{The same as figure \protect\ref%
{g3} but for the source mass consisting of 4 quarters}
\label{f2}
\end{figure}

Instead of equations (\ref{36}) - (\ref{40}) we arrived to the following
results 
\end{subequations}
\begin{subequations}
\label{S.7}
\begin{gather}
z_{1}=-0.244\text{m, }z_{2}=0.483\text{m,}v_{z1}=v_{z2}=v=2.377\text{m s$%
^{-1}$},  \label{S.7a} \\
\left\{ \phi _{s}^{\left( C\right) }\left( z_{1},v\right) ,\phi _{s}^{\left(
C\right) }\left( z_{2},v\right) \right\} =\left\{ 1.73227\text{rad,}-1.67980%
\text{rad}\right\} ,  \label{S.7b} \\
\left\{ h_{1},h_{2},h_{3},h_{4}\right\} =\left\{ -0.391\text{m},-0.136\text{m%
},0.738\text{m, }0.933\text{m}\right\} ,  \label{S.7c} \\
\left\{ \phi _{s}^{\left( F\right) }\left( z_{1},v\right) ,\phi _{s}^{\left(
F\right) }\left( z_{2},v\right) \right\} =\left\{ -0.58469\text{rad},0.72938%
\text{rad}\right\} ,  \label{S.7d} \\
\Delta ^{2}\phi =4.72618\text{rad.}  \label{S.7e}
\end{gather}%
Relative contributions to the PDD listed now in the table \ref{t5}. Using
data from this table and equations (\ref{6a}), (\ref{6b}), (\ref{10}), and (%
\ref{10.1}), we obtained following error budget and shift 
\begin{widetext}
\begin{table}[tbp]
\caption{The same as in table \protect\ref{t1} but for the source mass
geometry shown in figure \protect\ref{f2}.}%
\label{t5}%
\begin{tabular}{|c|c|c|}
\hline
Term & C-configuration & F-configuration \\ \hline
$\phi _{s}^{I}\left( z_{i},v_{zi}\right) /\Delta ^{2}\phi $ & $0.367,~-0.355$
& -$0.124,~0.154$ \\ \hline
Linear in position & $%
\begin{array}{c}
10^{-11}\left( 3.88\delta z_{1C}-16.5\delta z_{2C}\right) +2.01\cdot 10^{5}.
\\ 
\times \left[ \Gamma _{E31}\left( \delta x_{1C}-\delta x_{2C}\right) +\Gamma
_{E32}\left( \delta y_{1C}-\delta y_{2C}\right) \right]%
\end{array}%
$ & $%
\begin{array}{c}
10^{-9}\left[ -6.50\delta z_{1F}-7.40\delta z_{2F}\right] -2.01\cdot 10^{5}
\\ 
\times \left[ \Gamma _{E31}\left( \delta x_{1F}-\delta x_{2F}\right) +\Gamma
_{E32}\left( \delta y_{1F}-\delta y_{2F}\right) \right]%
\end{array}%
$ \\ \hline
Linear in velocity & $%
\begin{array}{c}
10^{-12}\left( 9.49\delta v_{z1C}-28.3\delta v_{z2C}\right) +4.89\cdot 10^{4}
\\ 
\times \left[ \Gamma _{E31}\left( \delta v_{x1C}-\delta v_{x2C}\right)
+\Gamma _{E32}\left( \delta v_{y1C}-\delta v_{y2C}\right) \right]%
\end{array}%
$ & $%
\begin{array}{c}
10^{-9}\left[ -1.58\delta v_{z1F}-1.76\delta v_{z2F}\right] -4.89\cdot 10^{4}
\\ 
\times \left[ \Gamma _{E31}\left( \delta v_{x1F}-\delta v_{x2F}\right)
+\Gamma _{E32}\left( \delta v_{y1F}-\delta v_{y2F}\right) \right]%
\end{array}%
$ \\ \hline
Nonlinear in position & $%
\begin{array}{c}
3.22\delta x_{1C}^{2}+3.21\delta y_{1C}^{2}-6.43\delta z_{1C}^{2} \\ 
+3.62\delta x_{2C}^{2}+3.61\delta y_{2C}^{2}-7.24\delta z_{2C}^{2}%
\end{array}%
$ & $%
\begin{array}{c}
3.45\left( \delta x_{1F}^{2}+\delta y_{1F}^{2}\right) -6.90\delta z_{1F}^{2}
\\ 
+3.80\left( \delta x_{2F}^{2}+\delta y_{2F}^{2}\right) -7.60\delta z_{2F}^{2}%
\end{array}%
$ \\ \hline
Nonlinear in velocity & $%
\begin{array}{c}
0.212\delta v_{x1C}^{2}+0.211\delta v_{y1C}^{2}-0.423\delta v_{z1C}^{2} \\ 
+0.223\left( \delta v_{x2C}^{2}+\delta v_{y2C}^{2}\right) -0.446\delta
v_{z2C}^{2}%
\end{array}%
$ & $%
\begin{array}{c}
0.209\left( \delta v_{x1F}^{2}+\delta v_{y1F}^{2}\right) -0.418\delta
v_{z1F}^{2} \\ 
+0.246\left( \delta v_{x2F}^{2}+\delta v_{y2F}^{2}\right) -0.492v_{z2F}^{2}%
\end{array}%
$ \\ \hline
$%
\begin{tabular}{c}
Position-velocity \\ 
cross term%
\end{tabular}%
\ $ & $%
\begin{array}{c}
1.56\left( \delta v_{x1C}\delta x_{1C}+\delta v_{y1C}\delta y_{1C}\right)
-3.12\delta v_{z1C}\delta z_{1C} \\ 
+1.76\left( \delta v_{x2C}\delta x_{2C}+\delta v_{y2C}\delta y_{2C}\right)
-3.52\delta v_{z2C}\delta z_{2C}%
\end{array}%
$ & $%
\begin{array}{c}
1.68\left( \delta v_{x1F}\delta x_{1F}+\delta v_{y1F}\delta y_{1F}\right)
-3.35\delta v_{z1F}\delta z_{1F} \\ 
+1.85\left( \delta v_{x2F}\delta x_{2F}+\delta v_{y2F}\delta y_{2F}\right)
-3.69\delta v_{z2F}\delta z_{2F}%
\end{array}%
$ \\ \hline
\end{tabular}%
\end{table}
\begin{subequations}
\label{S.8}
\begin{align}
\sigma \left( \Delta _{s}^{\left( 2\right) }\phi \right) & =\left\{
10^{9}\dsum_{j=1,2}\dsum_{I=C,F}\left[ 40.5\left( \Gamma _{E31}^{2}\sigma
^{2}\left( x_{jI}\right) +\Gamma _{E32}^{2}\sigma ^{2}\left( y_{jI}\right)
\right) +2.39\left( \Gamma _{E31}^{2}\sigma ^{2}\left( v_{xjI}\right)
+\Gamma _{E32}^{2}\sigma ^{2}\left( v_{yjI}\right) \right) \right] \right.  
\notag \\
& +20.7\sigma ^{4}\left( x_{1C}\right) +20.6\sigma ^{4}\left( y_{1C}\right)
+82.6\sigma ^{4}\left( z_{1C}\right) +26.3\sigma ^{4}\left( x_{2C}\right)
+26.1\sigma ^{4}\left( y_{2C}\right) +105\sigma ^{4}\left( z_{2C}\right)  
\notag \\
& +0.0896\sigma ^{4}\left( v_{x1C}\right) +0.0889\sigma ^{4}\left(
v_{y1C}\right) +0.357\sigma ^{4}\left( v_{z1C}\right) +0.0998\sigma
^{4}\left( v_{x2C}\right) +0.0991\sigma ^{4}\left( v_{y2C}\right)
+0.398\sigma ^{4}\left( v_{z2C}\right)   \notag \\
& +2.45\sigma ^{2}\left( x_{1C}\right) \sigma ^{2}\left( v_{x1C}\right)
+2.43\sigma ^{2}\left( y_{1C}\right) \sigma ^{2}\left( v_{y1C}\right)
+9.75\sigma ^{2}\left( z_{1C}\right) \sigma ^{2}\left( v_{z1C}\right)  
\notag \\
& +3.10\sigma ^{2}\left( x_{2C}\right) \sigma ^{2}\left( v_{x2C}\right)
+3.08\sigma ^{2}\left( y_{2C}\right) \sigma ^{2}\left( v_{y2C}\right)
+12.4\sigma ^{2}\left( z_{2C}\right) \sigma ^{2}\left( v_{z2C}\right)  
\notag \\
& +23.8\left[ \sigma ^{4}\left( x_{1F}\right) +\sigma ^{4}\left(
y_{1F}\right) \right] +95.2\sigma ^{4}\left( z_{1F}\right) +28.9\left[
\sigma ^{4}\left( x_{2F}\right) +\sigma ^{4}\left( y_{2F}\right) \right]
+116\sigma ^{4}\left( z_{2F}\right) +  \notag \\
& +0.0874\sigma ^{4}\left( v_{x1F}\right) +0.0872\sigma ^{4}\left(
v_{y1F}\right) +0.349\sigma ^{4}\left( v_{z1F}\right) +0.121\left[ \sigma
^{4}\left( v_{x2F}\right) +\sigma ^{4}\left( v_{y2F}\right) \right]
+0.485\sigma ^{4}\left( v_{z2F}\right)   \notag \\
& +2.81\left[ \sigma ^{2}\left( x_{1F}\right) \sigma ^{2}\left(
v_{x1F}\right) +\sigma ^{2}\left( y_{1F}\right) \sigma ^{2}\left(
v_{y1F}\right) \right] +11.2\sigma ^{2}\left( z_{1F}\right) \sigma
^{2}\left( v_{z1F}\right)   \notag \\
& \left. +3.42\sigma ^{2}\left( x_{2F}\right) \sigma ^{2}\left(
v_{x2F}\right) +3.41\sigma ^{2}\left( y_{2F}\right) \sigma ^{2}\left(
v_{y2F}\right) +13.7\sigma ^{2}\left( z_{2F}\right) \sigma ^{2}\left(
v_{z2F}\right) \right\} ^{1/2},  \label{S.8a} \\
s\left( \Delta _{s}^{\left( 2\right) }\phi \right) & =3.22\sigma ^{2}\left(
x_{1C}\right) +3.21\sigma ^{2}\left( y_{1C}\right) -6.43\sigma ^{2}\left(
z_{1C}\right) +3.62\sigma ^{2}\left( x_{2C}\right) +3.61\sigma ^{2}\left(
y_{2C}\right) -7.24\sigma ^{2}\left( v_{z2C}\right)   \notag \\
& +0.212\sigma ^{2}\left( v_{x1C}\right) +0.211\sigma ^{2}\left(
v_{y1C}\right) -0.423\sigma ^{2}\left( v_{z1C}\right) +0.223\left[ \sigma
^{2}\left( v_{x2C}\right) +\sigma ^{2}\left( v_{y2C}\right) \right]
-0.446\sigma ^{2}\left( v_{z2C}\right)   \notag \\
& +3.45\left[ \sigma ^{2}\left( x_{1F}\right) +\sigma ^{2}\left(
y_{1F}\right) \right] -6.90\sigma ^{2}\left( z_{1F}\right) +3.80\left[
\sigma ^{2}\left( x_{2F}\right) +\sigma ^{2}\left( y_{2F}\right) \right]
-7.60\sigma ^{2}\left( z_{2F}\right)   \notag \\
& +0.209\left[ \sigma ^{2}\left( v_{x1F}\right) +\sigma ^{2}\left(
v_{y1F}\right) \right] -0.418\sigma ^{2}\left( v_{z1F}\right) +0.246\left[
\sigma ^{2}\left( v_{x2F}\right) +\sigma ^{2}\left( v_{y2F}\right) \right]
-0.492\sigma ^{2}\left( v_{z2F}\right) .  \label{S.8b}
\end{align}
\end{subequations}
\end{widetext}Substituting here uncertainties of the atomic variables
equations (\ref{17a}), (\ref{17b}), and (\ref{35}) one gets instead of
equations (\ref{41}) and (\ref{42}) 
\end{subequations}
\begin{subequations}
\label{S.9}
\begin{align}
\sigma \left( \delta \Delta _{s}^{\left( 2\right) }\phi \right) & =17\text{%
ppm}\left[ 1+6.54\cdot 10^{14}\left( \Gamma _{E31}^{2}+\Gamma
_{E32}^{2}\right) \right] ^{1/2}\text{,}  \label{S.9a} \\
s\left( \delta \Delta _{s}^{\left( 2\right) }\phi \right) & =35\text{ppm,}
\label{S.9b} \\
\sqrt{\Gamma _{E31}^{2}+\Gamma _{E32}^{2}}& <18\text{E.}  \label{S.9c}
\end{align}

\section{Gravity field of the homogeneous cylinder}

\subsection{Axial component}

It is convenient \cite{c14} to explore the following expression for the
potential of the gravitational field of a homogeneous cylinder $\Phi \left( 
\mathbf{x}\right) $ 
\end{subequations}
\begin{equation}
\Phi \left( r,z\right) =-2G\rho \int_{0}^{R}dy\int_{r-\sqrt{R^{2}-y^{2}}}^{r+%
\sqrt{R^{2}-y^{2}}}d\xi \int_{z-h}^{z}\dfrac{d\zeta }{\sqrt{y^{2}+\xi
^{2}+\zeta ^{2}}},  \label{a1}
\end{equation}%
where $\rho ,$ $R,$ and $h$ are the density, radius, and height of the
cylinder, $\left( r,z,\psi =0\right) $ are the cylindrical coordinates of
the vector $\mathbf{x}$.. For an axial component of the gravitational field, 
$\delta g_{3}\left( r,z\right) =-\partial _{z}\Phi \left( r,z\right) $, one
gets%
\begin{equation}
\delta g_{3}\left( r,z\right) =2G\rho g_{3}\left( r,\zeta \right) _{\zeta
=z-h}^{\zeta =z},  \label{a2}
\end{equation}%
where the function 
\begin{equation}
g_{3}\left( r,\zeta \right) =\int_{0}^{R}dy\int_{r-\sqrt{R^{2}-y^{2}}}^{r+%
\sqrt{R^{2}-y^{2}}}\dfrac{d\xi }{\sqrt{y^{2}+\xi ^{2}+\zeta ^{2}}}
\label{a3}
\end{equation}%
can be represented as 
\begin{subequations}
\label{a4}
\begin{gather}
g_{3}\left( r,\zeta \right) =\int_{0}^{R}dy\ln \dfrac{t_{+}\left( y\right) }{%
t_{-}\left( y\right) }  \notag \\
=-\int_{0}^{R}y\left( \dfrac{dt_{+}}{t_{+}}-\dfrac{dt_{-}}{t_{-}}\right) ,
\label{a4a} \\
t_{\pm }\left( y\right) =r\pm \sqrt{R^{2}-y^{2}}  \notag \\
+\left( \zeta ^{2}+r^{2}+R^{2}\pm 2r\sqrt{R^{2}-y^{2}}\right) ^{1/2}
\label{a4b}
\end{gather}%
Since $t_{+}\left( R\right) =t_{-}\left( R\right) \equiv t\left( R\right)
\lessgtr t_{\pm }\left( 0\right) $ one can write 
\end{subequations}
\begin{equation}
g_{3}\left( r,\zeta \right) =\int_{t\left( R\right) }^{t_{+}\left( 0\right) }%
\dfrac{dt}{t}y_{+}\left( t\right) +\int_{t_{-}\left( 0\right) }^{t\left(
R\right) }\dfrac{dt}{t}y_{-}\left( t\right) ,  \label{a5}
\end{equation}%
where $y_{\pm }\left( t\right) $ is the root of the equation $t_{\pm }\left(
y\right) =t$. To find this root, consider the functions $x_{\pm }\left(
t\right) =\sqrt{R^{2}-y_{\pm }^{2}\left( t\right) },$%
\begin{equation}
0<x_{\pm }\left( t\right) <R.  \label{a6}
\end{equation}%
For them one gets%
\begin{equation}
x_{\pm }\left( t\right) =\pm t+\sqrt{\zeta ^{2}+R^{2}+2tr}\text{ or }\pm t-%
\sqrt{\zeta ^{2}+R^{2}+2tr}  \label{a7}
\end{equation}%
Since $t+\sqrt{\zeta ^{2}+R^{2}+2tr}>R$, then one should choose $x_{+}\left(
t\right) =t-\sqrt{\zeta ^{2}+R^{2}+2tr}.$ Since $t_{-}\left( 0\right)
>r-R+\left\vert r-R\right\vert >0,-t-\sqrt{\zeta ^{2}+R^{2}+2tr}<0$, hence $%
x_{-}\left( t\right) =\sqrt{\zeta ^{2}+R^{2}+2tr}-t$ or%
\begin{equation}
x_{\pm }\left( t\right) =\pm \left( t-\sqrt{\zeta ^{2}+R^{2}+2tr}\right) .
\label{a8}
\end{equation}%
Therefore, one concludes that the functions $y_{\pm }\left( t\right) $ are
coincident and equal to%
\begin{gather}
y_{+}\left( t\right) =y_{-}\left( t\right) =y\left( t\right) =  \notag \\
\left[ 2t\left( \sqrt{\zeta ^{2}+R^{2}+2tr}-r\right) -t^{2}-\zeta ^{2}\right]
^{1/2}  \label{a9}
\end{gather}%
and.%
\begin{equation}
g_{3}\left( r,\zeta \right) =\int_{t_{-}\left( 0\right) }^{t_{+}\left(
0\right) }\dfrac{dt}{t}y\left( t\right) .  \label{a10}
\end{equation}%
Introducing new variable,%
\begin{equation}
u=\sqrt{\zeta ^{2}+R^{2}+2tr}-r,  \label{a11}
\end{equation}%
for which 
\begin{subequations}
\label{a12}
\begin{eqnarray}
u\left[ t_{\pm }\left( 0\right) \right] &\equiv &u_{\pm }=\sqrt{\zeta
^{2}+\left( r\pm R\right) ^{2}},  \label{a12a} \\
y\left( t\right) &=&\dfrac{\sqrt{q\left( u^{2}\right) }}{2r},  \label{a12b}
\\
q\left( \eta \right) &=&\left\{ u_{+}^{2}-\eta \right\} \left\{ \eta
-u_{-}^{2}\right\} ,  \label{a12c} \\
dt &=&\dfrac{u+r}{r}du  \label{a12d}
\end{eqnarray}%
and so 
\end{subequations}
\begin{subequations}
\label{a13}
\begin{eqnarray}
g_{3}\left( r,\zeta \right) &=&I+I^{\prime },  \label{a13a} \\
I &=&\int_{u_{-}}^{u_{+}}\dfrac{du}{\sqrt{q\left( u^{2}\right) }}J\left(
u\right) ,  \label{a13b} \\
J\left( u\right) &=&\dfrac{q\left( u^{2}\right) \left( r^{2}-\zeta
^{2}-R^{2}-u^{2}\right) }{w\left( u^{2}\right) },  \label{a13c} \\
I^{\prime } &=&\dfrac{1}{2r}\int_{u_{-}^{2}}^{u_{+}^{2}}d\eta J^{\prime
}\left( \eta \right) ,  \label{a13d} \\
J^{\prime }\left( \eta \right) &=&\dfrac{\left( \eta -\zeta
^{2}-R^{2}-r^{2}\right) }{\sqrt{q\left( \eta \right) }w\left( \eta \right) },
\label{a13e} \\
w\left( \eta \right) &=&\left( \eta -\eta _{1}\right) \left( \eta -\eta
_{2}\right) ,  \label{a13f} \\
\eta _{1,2} &=&\left( r\pm \sqrt{\zeta ^{2}+R^{2}}\right) ^{2}.  \label{a13g}
\end{eqnarray}%
Using equality 
\end{subequations}
\begin{equation}
w\left( \eta \right) +q\left( \eta \right) =-4r^{2}\zeta ^{2},  \label{a14}
\end{equation}%
one can show that the integrand $J^{\prime }\left( \eta \right) $ is an
antisymmetric function with respect to the middle point $\eta =\left\{ u^{2}%
\left[ t_{+}\left( 0\right) \right] +u^{2}\left[ t_{-}\left( 0\right) \right]
\right\} /2,$ and, therefore, the term (\ref{a13d}) is equal $0$. At the
same time, expanding $J\left( u\right) $ into partial fractions, one obtains 
\begin{subequations}
\label{a15}
\begin{eqnarray}
g_{3}\left( r,\zeta \right) &=&\left( R^{2}+\zeta ^{2}-r^{2}\right)
I_{1}+I_{2}+I_{3+}+I_{3-},  \label{a15a} \\
I_{1} &=&\int_{u_{-}}^{u_{+}}\dfrac{du}{\sqrt{q\left( u^{2}\right) }},
\label{a15b} \\
I_{2} &=&\int_{u_{-}}^{u_{+}}\dfrac{duu^{2}}{\sqrt{q\left( u^{2}\right) }},
\label{a15c} \\
I_{3\pm } &=&2r\zeta ^{2}\left( r\pm \sqrt{\zeta ^{2}+R^{2}}\right)  \notag
\\
&&\times \int_{u_{-}}^{u_{+}}\dfrac{du}{\sqrt{q\left( u^{2}\right) }\left(
u^{2}-\eta _{1,2}\right) }  \label{a15d}
\end{eqnarray}%
The integrals (equations (\ref{a15})), one can compute using the
substitution 
\end{subequations}
\begin{equation}
u=\sqrt{u_{+}^{2}-\left( u_{+}^{2}-u_{-}^{2}\right) \sin ^{2}\phi }
\label{a15.1}
\end{equation}%
Finally, one arrives at the following expression for the axial component of\
the cylinder's field 
\begin{subequations}
\label{a16}
\begin{gather}
\delta g_{3}\left( r,z\right) =2G\rho g_{3}\left( r,\zeta \right) _{\zeta
=z-h}^{\zeta =z},  \label{a16a} \\
g_{3}\left( r,\zeta \right) =\dfrac{\left( \zeta ^{2}+R^{2}-r^{2}\right) }{%
\sqrt{\zeta ^{2}+\left( r+R\right) ^{2}}}K\left( k\right)  \notag \\
+\sqrt{\zeta ^{2}+\left( r+R\right) ^{2}}E\left( k\right) +\dfrac{\zeta ^{2}%
}{\sqrt{\zeta ^{2}+\left( r+R\right) ^{2}}}  \notag \\
\times \dsum_{j=\pm 1}\left[ \dfrac{r+j\sqrt{\zeta ^{2}+R^{2}}}{R-j\sqrt{%
\zeta ^{2}+R^{2}}}\Pi \left( \dfrac{2R}{R-j\sqrt{\zeta ^{2}+R^{2}}}|k\right) %
\right] ,  \label{a16b} \\
k=\sqrt{\dfrac{4rR}{\zeta ^{2}+\left( r+R\right) ^{2}}},  \label{a16c}
\end{gather}%
where $K\left( k\right) ,$ $E\left( k\right) $ and $\Pi \left( \alpha
|k\right) $ are the complete elliptic integrals of the first, second and
third order respectively.

\subsection{Radial component}

For the radial component of the gravitational field $\delta g_{r}\left(
r,z\right) =-\partial _{r}\Phi \left( r,z\right) $ one obtains from equation
(\ref{a1}) 
\end{subequations}
\begin{subequations}
\label{a17}
\begin{gather}
\delta g_{r}\left( r,z\right) =2G\rho g_{r}\left( r,\zeta \right) _{\zeta
=z-h}^{\zeta =z},  \label{a17a} \\
g_{r}\left( r,\zeta \right) =-\int_{0}^{R}y\left( \dfrac{dt_{+}}{t_{+}}-%
\dfrac{dt_{-}}{t_{-}}\right) ,  \label{a17b} \\
t_{\pm }\left( y\right) =\zeta +\left[ \zeta ^{2}+r^{2}+R^{2}\pm 2r\sqrt{%
R^{2}-y^{2}}\right] ^{1/2}  \label{a17c}
\end{gather}%
Since still $t_{+}\left( R\right) =t_{-}\left( R\right) \lessgtr t_{\pm
}\left( 0\right) ,$ one gets, 
\end{subequations}
\begin{equation}
g_{r}\left( r,\zeta \right) =\int_{t_{-}\left( 0\right) }^{t\left( R\right) }%
\dfrac{dt}{t}y_{-}\left( t\right) +\int_{t\left( R\right) }^{t_{+}\left(
0\right) }\dfrac{dt}{t}y_{+}\left( t\right) ,  \label{a17.1}
\end{equation}%
where $y_{\pm }\left( t\right) $ are functions inverse to equation (\ref%
{a17c}). Since these functions are the same%
\begin{eqnarray}
y_{+}\left( t\right) &=&y_{-}\left( t\right) \equiv y\left( t\right)  \notag
\\
&=&\dfrac{1}{2r}\left[ 4r^{2}R^{2}-\left( t^{2}-2\zeta t-r^{2}-R^{2}\right)
^{2}\right] ^{1/2},  \label{a18}
\end{eqnarray}%
then, choosing as an integration variable $u=t-\zeta $, one finds that 
\begin{subequations}
\label{a19}
\begin{eqnarray}
g_{r}\left( r,\zeta \right) &=&I+I^{\prime },  \label{a19a} \\
I &=&-\dfrac{\zeta }{2r}\int_{u_{-}}^{u_{+}}\dfrac{duq\left( u^{2}\right) }{%
\left( u^{2}-\zeta ^{2}\right) \sqrt{q\left( u^{2}\right) }},  \label{a19b}
\\
I^{\prime } &=&\dfrac{1}{4r}\int_{u_{-}^{2}}^{u_{+}^{2}}\dfrac{d\eta \sqrt{%
q\left( \eta \right) }}{\left( \eta -\zeta ^{2}\right) },  \label{a19c}
\end{eqnarray}%
where $u_{\pm }$ and $q\left( \eta \right) $ are given by equations (\ref%
{a12a}) and (\ref{a12c})$.$ Because $u_{\pm }^{2}-\zeta ^{2}$ and $q\left(
\eta +\zeta ^{2}\right) $ are independent of $\zeta $, the term $I^{\prime }$
gives no contribution to the acceleration (equation (\ref{a17a})) and can be
omitted. While using the substitution equation (\ref{a15.1}), one reduces
the integral in equation (\ref{a19b}) to elliptic integrals, which brings us
to the next final result

\end{subequations}
\begin{subequations}
\label{a20}
\begin{gather}
\delta g_{r}\left( r,z\right) =2G\rho g_{r}\left( r,\zeta \right) _{\zeta
=z-h}^{\zeta =z},  \label{a20a} \\
g_{r}\left( r,\zeta \right) =\dfrac{\zeta }{2r\sqrt{\zeta ^{2}+\left(
r+R\right) ^{2}}}  \notag \\
\times \left[ -\left( \zeta ^{2}+2r^{2}+2R^{2}\right) K\left( k\right)
+\left( \zeta ^{2}+\left( r+R\right) ^{2}\right) E\left( k\right) \right. 
\notag \\
\left. +\dfrac{\left( r^{2}-R^{2}\right) ^{2}}{\left( r+R\right) ^{2}}\Pi
\left( \dfrac{4rR}{\left( r+R\right) ^{2}}|k\right) \right] ,  \label{a20b}
\end{gather}%
where $k$ is given by equation (\ref{a16c}).

\end{subequations}

\end{document}